\documentclass[twocolumn]{aastex61}
\usepackage{url}
\usepackage{epsfig}
\shorttitle{Extreme nature of 4 BluDOGs revealed by optical spectroscopy}
\shortauthors{Noboriguchi et al.}

\begin{document}
\title{Extreme nature of four blue-excess dust-obscured galaxies revealed by optical spectroscopy}
\correspondingauthor{Akatoki Noboriguchi}
\email{akatoki@shinshu-u.ac.jp}

\author[0000-0002-5197-8944]{Akatoki Noboriguchi}
\affiliation{School of General Education, Shinshu University, 3-1-1 Asahi, Matsumoto, Nagano 390-8621, Japan}
\affiliation{Frontier Research Institute for Interdisciplinary Sciences, Tohoku University, 6-3 Aramaki, Aoba-ku, Sendai, Miyagi 980-8578, Japan}
\affiliation{Graduate School of Science and Engineering, Ehime University, 2-5 Bunkyo-cho, Matsuyama, Ehime 790-8577, Japan}
\author[0000-0002-7402-5441]{Tohru Nagao}
\affiliation{Research Center for Space and Cosmic Evolution, Ehime University, 2-5 Bunkyo-cho, Matsuyama, Ehime 790-8577, Japan}
\author[0000-0002-3531-7863]{Yoshiki Toba}
\affiliation{National Astronomical Observatory of Japan, 2-21-1 Osawa, Mitaka, Tokyo 181-8588, Japan}
\affiliation{Department of Astronomy, Kyoto University, Kitashirakawa-Oiwake-cho, Kyoto 606-8502, Japan}
\affiliation{Academia Sinica Institute of Astronomy and Astrophysics, 11F of Astronomy-Mathematics Building, AS/NTU, No.1, Section 4, Roosevelt Road, Taipei 10617, Taiwan}
\affiliation{Research Center for Space and Cosmic Evolution, Ehime University, 2-5 Bunkyo-cho, Matsuyama, Ehime 790-8577, Japan}
\author[0000-0002-4377-903X]{Kohei Ichikawa}
\affiliation{Frontier Research Institute for Interdisciplinary Sciences, Tohoku University, 6-3 Aramaki, Aoba-ku, Sendai, Miyagi 980-8578, Japan}
\author[0000-0002-1732-6387]{Masaru Kajisawa}
\affiliation{Graduate School of Science and Engineering, Ehime University, 2-5 Bunkyo-cho, Matsuyama, Ehime 790-8577, Japan}
\author{Nanako Kato}
\affiliation{Graduate School of Science and Engineering, Ehime University, 2-5 Bunkyo-cho, Matsuyama, Ehime 790-8577, Japan}
\author[0000-0002-3866-9645]{Toshihiro Kawaguchi}
\affiliation{Department of Economics, Management and Information Science Onomichi City University, Hisayamada 1600-2, Onomichi, Hiroshima 722-8506, Japan}
\author{Hideo Matsuhara}
\affiliation{Institute of Space and Astronautical Science, Japan Aerospace Exploration Agency, 3-1-1 Yoshinodai, Chuo-ku, Sagamihara, Kanagawa 252-5210, Japan}
\affiliation{Department of Space and Astronautical Science, The Graduate University for Advanced Studies, SOKENDAI, 3-1-1 Yoshinodai, Chuo-ku, Sagamihara, Kanagawa 252-5210, Japan}
\author[0000-0001-5063-0340]{Yoshiki Matsuoka}
\affiliation{Research Center for Space and Cosmic Evolution, Ehime University, 2-5 Bunkyo-cho, Matsuyama, Ehime 790-8577, Japan}
\author[0000-0002-0997-1060]{Kyoko Onishi}
\affiliation{Department of Space, Earth and Environment, Chalmers University of Technology, Onsala Observatory, SE-439 92 Onsala, Sweden}
\author[0000-0003-2984-6803]{Masafusa Onoue}
\affiliation{Kavli Institute for the Physics and Mathematics of the Universe (Kavli IPMU, WPI), The University of Tokyo, 5-1-5 Kashiwanoha, Kashiwa, Chiba 277-8583, Japan}
\affiliation{Kavli Institute for Astronomy and Astrophysics, Peking University, Beijing 100871, China}
\affiliation{Max-Planck-Institut f\"ur Astronomie, K\"onigstuhl 17, D-69117, Heidelberg, Germany}
\author{Nozomu Tamada}
\affiliation{Graduate School of Science and Engineering, Ehime University, 2-5 Bunkyo-cho, Matsuyama, Ehime 790-8577, Japan}
\author[0000-0001-5899-9185]{Koki Terao}
\affiliation{Subaru Telescope, National Astronomical Observatory of Japan, 650 North A'ohoku Place, Hilo, HI 96720, USA}
\affiliation{Astronomical Institute, Tohoku University, 6-3 Aramaki, Aoba-ku, Sendai, Miyagi 980-8578, Japan}
\author[0000-0003-1780-5481]{Yuichi Terashima}
\affiliation{Graduate School of Science and Engineering, Ehime University, 2-5 Bunkyo-cho, Matsuyama, Ehime 790-8577, Japan}
\affiliation{Research Center for Space and Cosmic Evolution, Ehime University, 2-5 Bunkyo-cho, Matsuyama, Ehime 790-8577, Japan}
\author[0000-0002-5485-2722]{Yoshihiro Ueda}
\affiliation{Department of Astronomy, Kyoto University, Kitashirakawa-Oiwake-cho, Kyoto 606-8502, Japan}
\author[0000-0002-4999-9965]{Takuji Yamashita}
\affiliation{National Astronomical Observatory of Japan, 2-21-1 Osawa, Mitaka, Tokyo 181-8588, Japan}



\begin{abstract}
We report optical spectroscopic observations of four blue-excess dust-obscured galaxies (BluDOGs) identified by Subaru Hyper Suprime-Cam. BluDOGs are a sub-class of dust-obscured galaxies (DOGs, defined with the extremely red color $(i-[22])_{\rm AB} \geq 7.0$; \citealt{2015PASJ...67...86T}), showing a significant flux excess in the optical $g$- and $r$-bands over the power-law fits to the fluxes at the longer wavelengths. \cite{2019ApJ...876..132N} has suggested that BluDOGs may correspond to the blowing-out phase involved in a gas-rich major merger scenario. However the detailed properties of BluDOGs are not understood because of the lack of spectroscopic information. In this work, we carry out deep optical spectroscopic observations of four BluDOGs using Subaru/FOCAS and VLT/FORS2. The obtained spectra show broad emission lines with extremely large equivalent widths, and a blue wing in the C~{\sc iv} line profile. The redshifts are between 2.2 and 3.3. The averaged rest-frame equivalent widths of the C~{\sc iv} lines are $160\pm33$ {\AA}, $\sim$7 times higher than the average of a typical type-1 quasar. The FWHMs of their velocity profiles are between 1990 and 4470 ${\rm km\ s^{-1}}$, and their asymmetric parameters are 0.05 and 0.25. Such strong C~{\sc iv} lines significantly affect the broad-band magnitudes, which is partly the origin of the blue excess seen in the spectral energy distribution of BluDOGs. Their estimated supermassive black hole masses are $1.1\times10^8 < M_{\rm BH}/M_\odot < 5.5 \times 10^8$. The inferred Eddington ratios of the BluDOGs are higher than 1 ($1.1< \lambda_{\rm Edd} < 3.8$), suggesting that the BluDOGs are in a rapidly evolving phase of supermassive black holes.

\end{abstract}



\section{Introduction} \label{sec:intro}
In the last two decades, observations of low-redshift galaxies have revealed tight correlations between the mass of supermassive black holes (SMBHs) and the host galaxy properties such as bulge mass (e.g., \citealt{1998AJ....115.2285M, 2000ApJ...539L...9F, 2000ApJ...539L..13G, 2002ApJ...574..740T, 2003ApJ...589L..21M, 2013ARA&A..51..511K, 2020ApJ...888...37D}). 
Such scaling relations suggest the so-called co-evolution between galaxies and SMBHs. 
It has been argued that a major merger of gas-rich galaxies triggers active star-forming activity and subsequent mass accretion onto SMBHs (e.g., \citealt{1988ApJ...325...74S, 2008ApJS..175..356H, 2012ApJ...758L..39T, 2018PASJ...70S..37G}).
In this scenario, the merging two galaxies first evolve into a dusty star-forming (SF) galaxy.
Then it evolves into a dusty active galactic nucleus (AGN) as gas accretion to the nuclear region triggers the activity of SMBHs.
Finally, a dusty AGN evolves into an optically-thin quasar after the surrounding dust is blown out by the powerful AGN outflow.
The most active period of such SF and AGN activity is generally heavily obscured by dust, which prevents us from investigating these phases observationally.

By combining optical, near-infrared (NIR), and mid-infrared (MIR) catalogs obtained from the Subaru Hyper Suprime-Cam (HSC; \citealt{2018PASJ...70S...1M})-Subaru Strategic Program (SSP; \citealt{2018PASJ...70S...4A}), the VISTA Kilo-degree Infrared Galaxy survey (VIKING; \citealt{2007Msngr.127...28A}), and the {\it Wide-field Infrared Survey Explorer} ({\it WISE}; \citealt{2010AJ....140.1868W}) all-sky survey (ALLWISE; \citealt{2014yCat.2328....0C}), \cite{2015PASJ...67...86T, 2017ApJ...835...36T} and \cite{2019ApJ...876..132N} selected dusty SF galaxies and/or powerful AGNs as dust-obscured galaxies (DOGs; \citealt{2008ApJ...677..943D, 2008ApJ...672...94F, 2009ApJ...705..184B, 2009ApJ...700.1190D, 2011ApJ...733...21B}).
DOGs are defined with a very red optical-MIR color ($(i-[22])_{\rm AB} \geq 7.0$; \citealt{2015PASJ...67...86T}).
DOGs represent a transition phase from a gas-rich major merger to an optically-thin quasar in the gas-rich major merger scenario (\citealt{2008ApJ...677..943D}), suggesting that some DOGs are expected to have buried AGNs.
Recently, eight blue-excess DOGs (BluDOGs; \citealt{2019ApJ...876..132N}) were discovered from the HSC-selected DOGs based on their optical spectral slopes (i.e., $\alpha_{\rm opt} < 0.4$, where  $\alpha_{\rm opt}$ is the observed-frame optical spectral index for the HSC $g$-, $r$-, $i$-, $z$-, and $y$-bands in the power-law fit, $f_{\nu}\propto\lambda^{\alpha_{\rm opt}}$), and are a very rare population (eight BluDOGs out of 571 HSC-selected DOGs).
\cite{2019ApJ...876..132N} suggested that the BluDOGs with such blue excess may be in the blowing-out phase involved in the gas-rich major merger scenario.
However, the detailed properties of BluDOGs are not well understood because of the lack of spectroscopic information. Spectroscopic observations will give us accurate redshifts, and thus reliable AGN luminosities as a measure of the accretion rates, as well as the SMBH masses.

Another interesting population that may represent the transition phase between optically-thick AGNs and optically-thin quasars is extremely red quasars (ERQs; e.g., \citealt{2015MNRAS.453.3932R, 2017MNRAS.464.3431H, 2019MNRAS.488.4126P, 2020A&A...634A.116V}).
ERQs were identified by combining the optical photometric data of Sloan Digital Sky Survey (SDSS; \citealt{2000AJ....120.1579Y}), the optical spectroscopic data from SDSS-III (\citealt{2011AJ....142...72E}) Baryon Oscillation Spectroscopic Survey (BOSS; \citealt{2013AJ....145...10D}), and the MIR photometric data of {\it WISE} catalog.
They are also defined with very red optical to MIR colors ($F_\nu(24{\rm \mu m})/ F_\nu(R) \geq 1000$), and their spectra show  broad emission lines with extremely large equivalent widths (\citealt{2015MNRAS.453.3932R, 2017MNRAS.464.3431H}).
\cite{2017MNRAS.464.3431H} refined the definition of ERQs as $(i-[12])_{\rm AB} > 4.6${\footnote{All of the BluDOGs also satisfy the criterion of the ERQ (see Table~\ref{tab:photodata}).}},
and reported notable blue-wing features in their C~{\sc iv} profiles, which suggests the presence of powerful outflow.
However, the ERQ sample is limited to optically bright objects since their selection requires SDSS spectra. Detailed studies of optically-faint populations in the transition phase between the optically-thick and optically-thin stages are required to understand the whole scenario of the merger-driven evolution of SMBHs.
Therefore, it is important to execute the spectroscopic observations for BluDOGs and to research their spectroscopic properties.

In this work, we present the results of spectroscopic observations and subsequent analyses of four BluDOGs.
This paper is organized as follows.
We describe sample selection of our targets and observations in Section {\ref{sec:data}}.
In Section {\ref{sec:results}}, we present properties of the detected emission lines, the estimated dust extinctions, bolometric luminosities of an AGN ($L^{\rm AGN}_{\rm bol}$), and SMBH masses ($M_{\rm BH}$).
The discussion on the large equivalent widths of the C~{\sc iv} emission, their SMBH mass, and Eddington ratios is given in Section {\ref{sec:disscussion}}.
Then we give a brief summary in Section {\ref{sec:conclusion}}.
Throughout this paper, the adopted cosmology is a flat universe with $H_0 = 70\ {\rm km\ s^{-1}\ Mpc^{-1}}$, $\Omega_M = 0.3$, and $\Omega_\Lambda = 0.7$.
Unless otherwise noted, all magnitudes refer to the AB system.

%%%%%%%%%%%%%%%%%%%%%%%%%
\begin{deluxetable*}{lrlll}
\tablecaption{Photometric data of BluDOGs\label{tab:photodata}}
\tablehead{
\colhead{Name}				& \colhead{HSC $r$-band} & \colhead{HSC $i$-band}	& \colhead{WISE $W3$-band} 	& \colhead{WISE $W4$-band}\\
\colhead{}						& \colhead{[AB mag]} 	& \colhead{[AB mag]} 		& \colhead{[AB mag]} 		& \colhead{[AB mag]}
}
\startdata
	HSC J090705.64$+$020955.8 (HSC J0907)	& 22.56$\pm$0.01 	& 22.59$\pm$0.01 	& 16.06$\pm$0.13 		& 14.89$\pm$0.34 	\\\hline
	HSC J120200.84$-$011846.4 (HSC J1202)	& 20.92$\pm$0.00 	& 20.87$\pm$0.00 	& 14.47$\pm$0.04 		& 13.46$\pm$0.10 	\\\hline
	HSC J120728.71$-$005808.4 (HSC J1207)	& 22.12$\pm$0.01 	& 22.31$\pm$0.01 	& 16.28$\pm$0.16 		& 15.01$\pm$0.36 	\\\hline
	HSC J141435.21$+$003547.4				& 23.32$\pm$0.02 	& 23.11$\pm$0.02 	& 17.24\tablenotemark{a} 	& 15.33$\pm$0.33 	\\\hline
	HSC J143727.40$-$011726.5				& 23.17$\pm$0.02 	& 23.10$\pm$0.01 	& 16.94$\pm$0.23 		& 15.37$\pm$0.31 	\\\hline
	HSC J144333.84$-$000830.3 (HSC J1443)	& 22.34$\pm$0.01 	& 22.24$\pm$0.01 	& 16.14$\pm$0.10 		& 15.04$\pm$0.23 	\\\hline
	HSC J144813.65$+$002244.3				& 23.55$\pm$0.02 	& 23.43$\pm$0.02 	& 16.83$\pm$0.16 		& 15.37$\pm$0.34 	\\\hline
	HSC J144900.84$+$002350.2				& 23.95$\pm$0.03 	& 23.74$\pm$0.02 	& 17.15$\pm$0.22 		& 15.47$\pm$0.36 	\\\hline
\enddata
\tablenotetext{a}{The magnitude is a 95\% confidence upper limit.\\\url{https://wise2.ipac.caltech.edu/docs/release/allwise/expsup/sec2_1a.html}}
\end{deluxetable*}

\begin{deluxetable*}{lrlll}
\tablecaption{Observation log\label{tab:status}}
\tablehead{
\colhead{Name}				& \colhead{Exp. time [s]} 	& \colhead{Date}		& \colhead{Standard star} 	& \colhead{Instrument}
}
\startdata
	HSC J0907 	& 900$\times$2 		& 2019 October 8 		& G191-B2B 			& FOCAS (Subaru) 	\\
						 			& 600$\times$1 		& 					& 		 			& 			 	\\\hline
	HSC J1202 	& 900$\times$6 		& 2019 February 27 		& LTT 6248 			& FORS2 (VLT) 	\\\hline
	HSC J1207 	& 900$\times$12 		& 2019 March 1, 2, 6 	& LTT 4816 			& FORS2 (VLT)  	\\\hline
	HSC J1443 	& 900$\times$12 		& 2019 March 7, 8 		& LTT 4816, EG 274 	& FORS2 (VLT) 	\\ \hline
\enddata
\end{deluxetable*}

\section{Sample and the data} \label{sec:data}

\subsection{Sample selection}

In \cite{2019ApJ...876..132N}, 571 DOGs were selected by combining $\sim$105 deg$^2$ imaging data obtained from the survey of HSC-SSP\footnote{ We utilize the photometric data of S16A HSC-SSP, which was released internally within the HSC survey team and is based on data obtained from 2014 March to 2016 April.} ($g$, $r$, $i$, $z$, and $y$), VIKING ($Z$, $Y$, $J$, $H$, and $Ks$), and ALLWISE ($W1$, $W2$, $W3$, and $W4$).
The eight BluDOGs were defined among the DOG sample with the smallest observed-frame optical slope  ($\alpha_{\rm opt } \textless 0.4$, where $\alpha_{\rm opt }$ is the observed-frame optical spectral index of the power-law fits to the HSC $g$, $r$, $i$, $z$, and $y$-band fluxes, $f_{\nu} \propto \lambda^{\alpha_{\rm opt}}$).
We selected the four brightest BluDOGs ($r_{\rm AB} < 23$: see Table~\ref{tab:photodata}) as the targets of our spectroscopic observations presented in this paper.

\subsection{Spectroscopic observations and data reductions}\label{subsec:so}

We executed the observations by using Faint Object Camera and Spectrograph (FOCAS; \citealt{2002PASJ...54..819K}) installed on the Subaru Telescope of National Astronomical Observatory of Japan, and FORS2 (\citealt{1998Msngr..94....1A}) installed on Very Large Telescope (VLT-UT1) of European Southern Observatory (ESO).
We present the observation log in Table~\ref{tab:status}.

\subsubsection{Subaru FOCAS}
By using FOCAS, we observed HSC J090705.64$+$020955.8 (hereafter J0907) on October 8th in 2019, with airmass $\sim$1.76 and seeing $\sim$0.5 arcsec.
We used the 300B grism and the SY47 filter to cover $\lambda_{\rm obs}\sim$4700--9200 {\AA}, with the resultant spectral resolution of $R \sim$800 for the used 0\arcsec.8-width slit.
To reduce the obtained data, we performed bias correction, flat fielding with dome flat, removal of cosmic-rays, spectral extraction, sky subtraction, wavelength calibration, and flux calibration with a standard star (G191-B2B) using the Python packages of \verb|Astropy| and \verb|Numpy|.
For removing cosmic-rays, we utilized \verb|Astro-SCRAPPY| (\citealt{2019ascl.soft07032M}).
\verb|Astro-SCRAPPY| is based on the algorithm of \verb|L.A.Cosmic|, which removes cosmic-rays based on a variation of Laplacian edge detection (\citealt{2001PASP..113.1420V}).
The final spectrum is an inverse-variance weighted mean of the individual shots, corrected for the Galactic extinction (\citealt{1998ApJ...500..525S}).

\subsubsection{VLT FORS2}
By using FORS2, we observed HSC J120200.84$-$011846.4, HSC J120728.71$-$005808.4, and HSC J144333.84$-$000830.3 (hereafter J1202, J1207, and J1443, respectively) between February 27th and March 8th, 2019.
We used the GRISM\_600RI$+19$ and the GG435 filter to cover $\lambda_{\rm obs}\sim$5200--8000 {\AA}, which results in the spectral resolution of $R\sim$1500 with 0\arcsec.7-width slit.
The typical airmasses of the observations for J1202, J1207, and J1443 were 1.17, 1.24, and 1.12, and the typical seeing sizes were $\sim$1.0, 0.5, and 0.5 arcsec, respectively.
For the data reduction, we utilized the \verb|Recipe flexible execution workbench| (\verb|Reflex|; \citealt{2013A&A...559A..96F}) software.
\verb|Reflex| performed bias correction, flat fielding with dome flat, sky subtraction, removing cosmic-rays, spectral extraction, wavelength calibration, and flux calibration with a standard star (LTT\ 6248, LTT\ 4816, and EG\ 274).
The final spectrum of each target is the inverse-variance weighted mean of the individual shots, corrected for the Galactic extinction.

\begin{figure*}%[ht!]
	\centering
	\vspace{-0.5cm}
	\includegraphics[width=18cm]{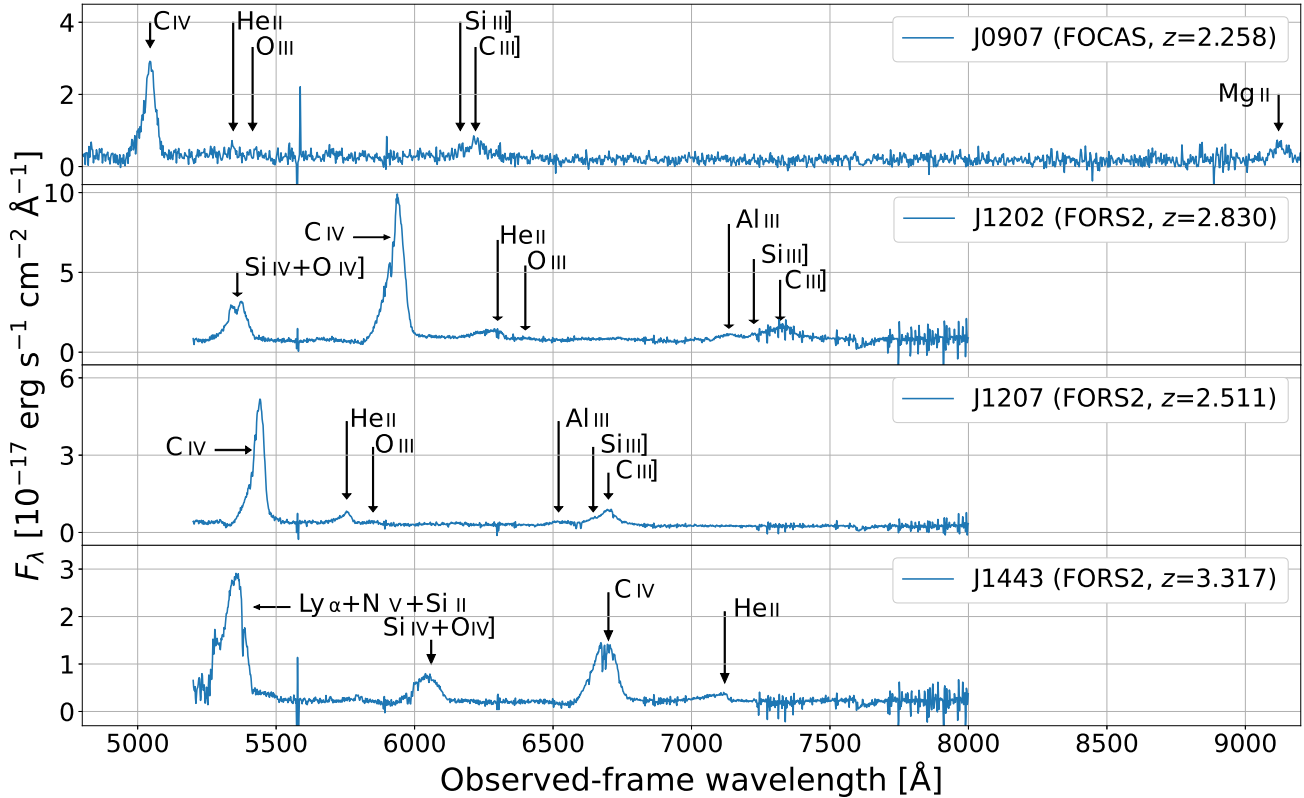} 
	\vspace{-1.0cm}
	\caption{The reduced spectra of the BluDOGs. The spectra are for J0907, J1202, J1207 and J1443 from the top to bottom. Detected lines are marked by arrows and labels.}
	\label{fig:bludogsp}
	%\vspace{1cm}
\end{figure*}

\subsubsection{Spectrophotometric re-calibration}
We re-calibrated the reduced spectra to match the HSC photometry, in order to correct for the effects of the slit loss of the flux, systematic errors in the photometric and spectroscopic calibrations, and any other possible systematic errors.
In our observations, the spectra cover the wavelength range of the HSC $r$-band.
We calculate the calibration factor, $f_{\rm photo\_calib} = F_{{\rm photo}\_r}/F_{{\rm spec}\_r}$, where $F_{{\rm photo}\_r}$ and $F_{{\rm spec}\_r}$ are the photometric and spectroscopic fluxes in the HSC $r$-band.
The derived calibration factors of J0907, J1202, J1207, and J1443 are 0.97, 1.50, 1.40, and 1.36, respectively.
We multiply the spectra with the derived calibration factors.

%%%%%%%%%%%%%%%%%%%%%%%%%%%%

\section{Results} \label{sec:results}

\begin{figure*}[ht]
	%\centering
	\vspace{-1cm}
	\includegraphics[width=18cm]{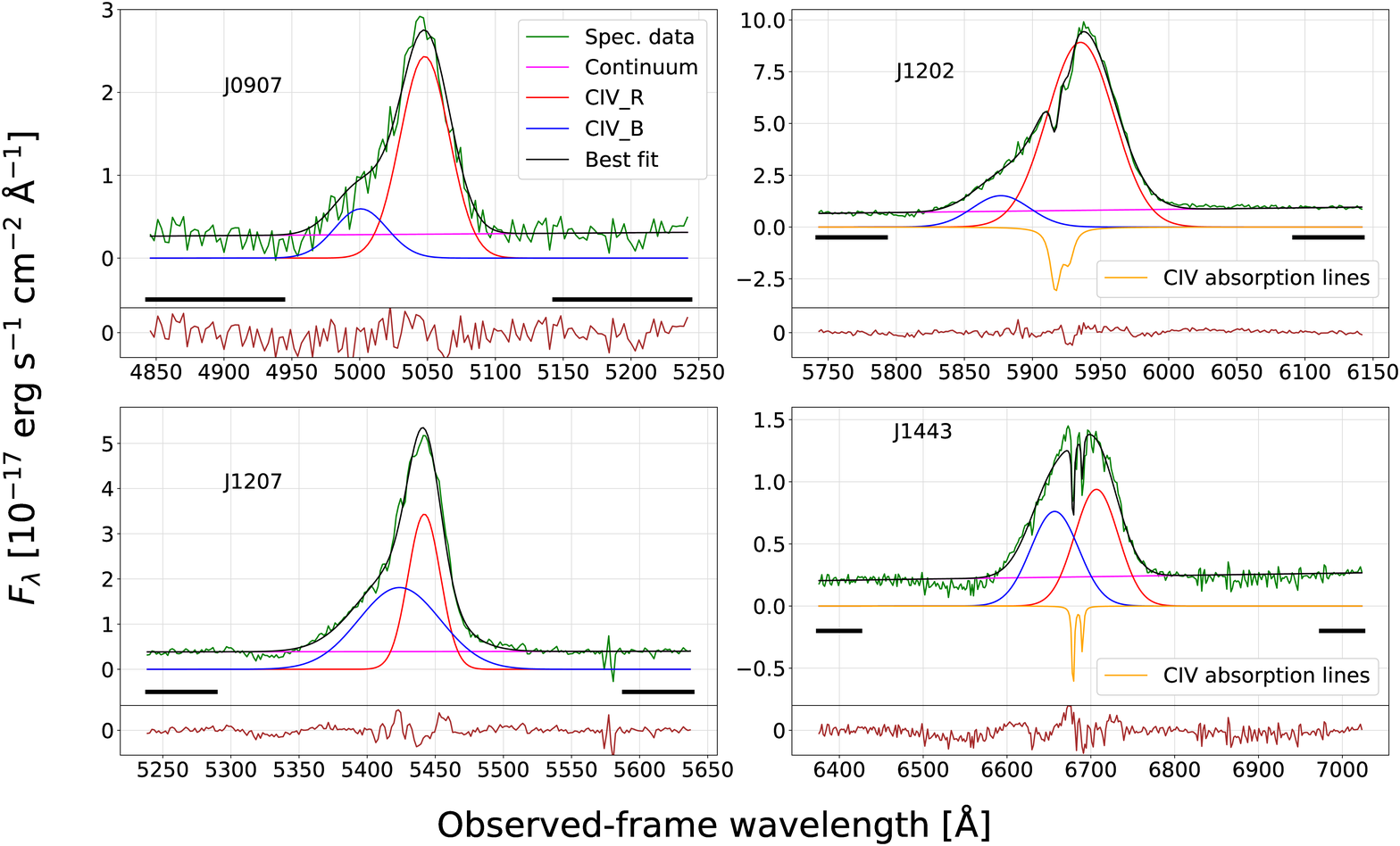} 
	\vspace{-1cm}
	%\vspace{30pt}
	\caption{Spectral fits to the C~{\sc iv} emission lines of the BluDOGs. The top left, top right, bottom left, and bottom right show the C~{\sc iv} emission lines of J0907, J1202, J1207, and J1443, respectively. The green, magenta, red, blue, and black lines represent the observed spectrum, linear fit to the continuum emission, two Gaussians for the red and blue components, and best-fit model, respectively. The orange line on the J1202 and J1443 panels represents the C~{\sc iv} doublet absorption line. The horizontal black bars denote the wavelength range used to fit the continuum emission. In each panel, the lower part presents the residual of the best-fit, with the same flux scale as in the upper part.\label{fig:BluCIV}}
\end{figure*}

%\vspace{-0.5cm}
\subsection{Emission-line measurements}\label{subsec:del}

Figure \ref{fig:bludogsp} shows the reduced spectra of the four BluDOGs. 
In order to measure the emission-line properties, we divide emission lines into six groups as follows; (1) Ly$\alpha$1216, N~{\sc v}$\lambda$1240, and Si~{\sc ii}$\lambda$1263, (2) Si~{\sc iv}$\lambda$1397 and O~{\sc iv}]$\lambda$1402, (3) He~{\sc ii}$\lambda$1640 and O~{\sc iii}]$\lambda$1663, and (4) Al~{\sc iii}$\lambda$1857, Si~{\sc iii}]$\lambda$1892, and C~{\sc iii}]$\lambda$1909, (5) C~{\sc iv} 1549, and (6) Mg~{\sc ii}. 
We fit the emission lines in each group simultaneously, with a linear continuum model subtracted from the observed spectrum. 
We adopt a single-Gaussian profile for Ly~${\alpha}$, N~{\sc v}, Si~{\sc ii}, Si~{\sc iv}, O~{\sc iv}], O~{\sc iii}], Al~{\sc iii}, Si~{\sc iii} and Mg~{\sc ii}.
The C~{\sc iii}] of J1202 is fitted with a single-Gaussian profile, while those of J0907 and J1207 are fitted with a double-Gaussian profile.
For the fit around the Si~{\sc iv} and O~{\sc iv}] of J1202, we add an additional Gaussian profile to reproduce the observed broad component.
We fit C~{\sc iv} and He~{\sc ii} with double-Gaussian profiles, and denote the blue and red components with the suffixes of  ``\_B'' and ``\_R'', respectively.
Additionally, we fit the doublet absorption lines observed around the C~{\sc iv} emission lines of J1202 and J1443.
The C~{\sc iv} absorption lines observed at $\lambda_{\rm obs}$ = 5916.8 {\AA} and 5926.6 {\AA} for J1202 and those at $\lambda_{\rm obs}$ = 6678.8 {\AA} and 6689.9 {\AA} for J1443 are fitted using the Voigt profile, respectively. 
The doublet absorption line ratio is fixed as 2:1 \citep{1983A&A...122..335F}.
The best values and standard deviations for emission and absorption lines parameters are estimated by using {\tt scipy.optimize.curve\_fit}\footnote{https://docs.scipy.org/doc/scipy/reference/}, while
we calculate full width at half maximum (FWHM) of emission lines with double Gaussian by using a Monte Carlo method.
For this Monte Carlo simulation, we created 10,000 mock spectra using the noise arrays of the observed spectra,
and calculate the mean and standard deviation of the line properties.
The results for emission lines are listed in Tables \ref{tab:dlineJ0907}--\ref{tab:dlineJ1443}.
For absorption lines, the observed-frame equivalent widths and redshifts of the doublet absorption lines on the J1202 C~{\sc iv} emission line are 47.7$\pm$19.6 {\AA}, 23.6$\pm$9.7 {\AA}, and 2.822, respectively, while the observed-frame equivalent widths and redshift of the doublet absorption lines on J1443 C~{\sc iv} emission line are 14.0$\pm$6.3 {\AA}, 6.94$\pm$3.14 {\AA}, and 3.314, respectively.
Therefore, the co-moving distance between J1202 and its C~{\sc iv} absorber is 8.73 Mpc, while that between J1443 and its C~{\sc iv} absorber is 2.79 Mpc.

The flux ratios of N~{\sc v}/Ly$\alpha$ and N~{\sc v}/C~{\sc iv} for J1443 are 3.9 and 1.8, respectively, whereas the values for the typical quasar \citep{2001AJ....122..549V} are 0.02 and 0.10.
One possible reason of these unusual flux ratios in J1443 is the presence of absorption lines, which absorb most of the Ly$\alpha$ and the C~{\sc iv} fluxes around the peak.
The unusual flux ratios cannot be explained by the dust reddening, given too small wavelength separations among emission lines of Ly$\alpha$, N~{\sc v}, and C~{\sc iv}.

Figure \ref{fig:BluCIV} showcases the best-fit models to the C~{\sc iv} emission lines in the four BluDOGs.
We adopt the C~{\sc iv} redshift taking C~{\sc iv}\_R $+$ C~{\sc iv}\_B into account as the systemic redshift of the targets.
The determined systemic redshifts of J0907, J1202, J1207, and J1443 are 2.258$\pm$0.002, 2.830$\pm$0.002, 2.511$\pm$0.001, and 3.317$\pm$0.006, respectively.

\subsection{Emission-line contributions to the HSC $g$- and $r$-band magnitudes}\label{subsec:EEWGR}

Figure~\ref{fig:bludogsp} suggests the very large equivalent width (EW) of the emission lines.
The average rest-frame EW (REW) of the C~{\sc iv} line of the four BluDOGs is $160\pm33$ {\AA}, $\sim$7 times higher than the average of SDSS type-1 quasars ($23.8\pm0.1$ {\AA}; \citealt{2001AJ....122..549V}). 
Here we investigate the effect of the large REWs  on the HSC $g$- and $r$-band magnitudes.

\begin{figure}%[ht!]
	\centering
	        %\hspace{-0.8cm}
		\includegraphics[width=8.5cm]{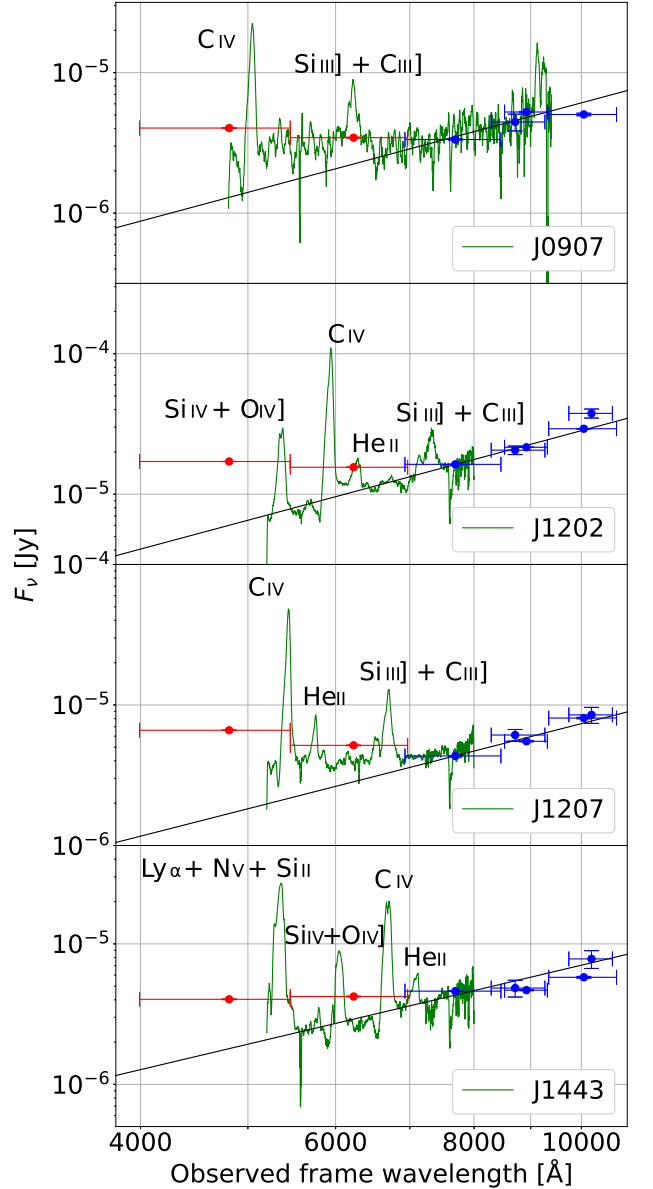} 
		%\vspace{30pt}
		\caption{The SED of J0907 (top), J1202 (middle upper), J1207 (middle lower), and J1443 (bottom). The red dots denote the $g$- and $r$-band magnitudes, while the blue dots denote the longer-wavelength optical and near-infrared magnitudes that are used for the power-law fit (black line). The green lines represent the observed spectra.}
		\label{fig:AFT}
\end{figure}

First, we calculate the expected magnitudes at the $g$- and $r$-bands from an extrapolation of the power-law fit to the longer wavelength bands ($i$, $z$, $y$, $Z$, $Y$, $J$, $H$, $Ks$, $W1$, $W2$, $W3$, and $W4$).
Figure~\ref{fig:AFT} clearly shows that the observed $g$- and $r$-band magnitudes exceed the extrapolation of the power-law fit.
The excesses of the $g$-band magnitudes for J0907, J1202, J1207, and J1443 are 1.27, 1.13, 1.48, and 0.88 mag, and those for the $r$-band excesses are 0.47, 0.44, 0.65, and 0.40 mag, respectively.

Furthermore, we estimate the effect of the strong emission lines, based on their observed-frame EWs and the band widths (BW) of the HSC $g$ and $r$-bands.
The BWs of the HSC $g$ and $r$-bands are 1468 and 1508 \AA\ \citep{2018PASJ...70...66K}, respectively.
By taking all of the emission lines (Tables \ref{tab:dlineJ0907} -- \ref{tab:dlineJ1443}) covered by the HSC $g$-band (4000--5500\AA) and $r$-band (5500--7000\AA) into account, the total observed-frame EWs for J0907, J1202, J1207, and J1443 in the $g$-band are 604, 258, 608, and 1380 \AA\  respectively, while those in the $r$-band are 129, 836, 329, and 721 {\AA} (Table~\ref{tab:EVEWS}).
Note that the total observed-frame EWs in the $g$-band are lower limits, because our optical spectra do not cover the entire wavelength range of the band (Section \ref{subsec:so}) and thus some emission lines are not taken into account in the derived total observed-frame EWs. 
Especially Ly$\alpha$, the strongest emission line in the rest-frame UV spectrum of typical AGNs, is not covered in our spectra of J0907, J1202, and J1207, thus the total observed-frame EWs for these 3 objects are largely underestimated\footnote{The Ly$\alpha$ line of J0907 is at the shorter edge of the HSC $g$-band coverage but the flux contribution to the $g$-band magnitude is likely to be significant owing to its broad nature.}.
Since the magnitude excess by the emission lines is given by $\Delta {\rm mag} = 2.5\log{(1+EW/BW)}$, the estimated effects in the $g$-band for J0907, J1202, J1207, and J1443 are 0.37, 0.18, 0.38, and 0.72 mag, respectively.
Similarly, the estimated effects of emission lines to the $r$-band magnitudes are 0.09, 0.48, 0.21, and 0.42 mag, respectively (see Table \ref{tab:EVEWS} for a summary). We will discuss the implication from these estimates in Section \ref{subsec:CIVEW}.

%%%%%%%%%%%%%%% Detected lines %%%%%%%%%%%%%%%%
%
%
%J0907
\begin{deluxetable*}{crrrrrr}
%\rotate
\tablecaption{The detected lines of J0907\label{tab:dlineJ0907}}
\tablehead{
\colhead{Line name} & \colhead{$\lambda_{\rm rest}$ [\AA]} & \colhead{$z_{\rm line}$} & \colhead{$FWHM_{\rm rest}$ [\AA]} & \colhead{$F_{\rm line}$ [${\rm erg\ s^{-1}\ cm^{-2}}$]}  & \colhead{$EW_{\rm rest}$ [\AA]} & \colhead{$v_{\rm width}$ [${\rm km\ s^{-1}}$]} \\
\colhead{(1)} & \colhead{(2)} & \colhead{(3)} & \colhead{(4)} & \colhead{(5)} & \colhead{(6)} & \colhead{(7)} 
}
\startdata
C~{\sc iv}\_R 					& 1549.5 & 2.258$\pm$0.001 & 12.8$\pm$0.7 		& (1.10$\pm$0.07)E$-$15 	& 118$\pm$10 		& 2470$\pm$130\\
C~{\sc iv}\_B 					& 1549.5 & 2.227$\pm$0.004 & 13.8$\pm$2.5 		& (2.87$\pm$0.60)E$-$16 	& 31.5$\pm$6.7 	& 2670$\pm$490\\
C~{\sc iv}\_R + C~{\sc iv}\_B 		& 1549.5 & 2.258$\pm$0.002 & 15.2$\pm$0.8 		& (1.39$\pm$0.09)E$-$15 	& 148$\pm$12 		& 2940$\pm$150\\\hline
He~{\sc ii}\_R 					& 1640.4 & 2.260$\pm$0.002 & 3.63$\pm$2.32 	& (3.50$\pm$2.36)E$-$17 	& 4.43$\pm$3.01 	& 663$\pm$425\\
He~{\sc ii}\_B 					& 1640.4 & 2.235$\pm$0.005 & 38.1$\pm$5.9 		& (2.13$\pm$0.48)E$-$16 	& 26.8$\pm$6.3 	& 6960$\pm$1080\\
He~{\sc ii}\_R + He~{\sc ii}\_B 		& 1640.4 & 2.260$\pm$0.002 & 5.54$\pm$1.29 	& (2.48$\pm$0.53)E$-$16 	& 31.4$\pm$7.0 	& 1010$\pm$240\\
O~{\sc iii}{]} 					& 1663.5 & 2.264$\pm$0.001 & 3.29$\pm$1.69 	& (4.49$\pm$1.48)E$-$17 	& 5.84$\pm$1.98 	& 593$\pm$305\\\hline
Si~{\sc iii} 						& 1892.0 & 2.258$\pm$0.002 & 4.03$\pm$4.16 	& (2.97$\pm$2.55)E$-$17 	& 3.56$\pm$3.06 	& 639$\pm$659\\
C~{\sc iii}{]}\_R 					& 1908.7 & 2.261$\pm$0.005 & 30.3$\pm$5.3 		& (2.34$\pm$0.66)E$-$16 	& 28.2$\pm$8.1 	& 4760$\pm$830\\
C~{\sc iii}{]}\_R 					& 1908.7 & 2.258$\pm$0.002 & 6.19$\pm$3.04 	& (6.50$\pm$3.21)E$-$17 	& 7.83$\pm$3.89 	& 971$\pm$477\\
C~{\sc iii}{]}\_R + C~{\sc iii}{]}\_B 	& 1908.7 & 2.258$\pm$0.003 & 11.6$\pm$1.7 		& (2.99$\pm$0.73)E$-$16 	& 36.0$\pm$9.0 	& 1830$\pm$260\\\hline
Mg~{\sc ii} 					& 2799.1 & 2.259$\pm$0.001 & 17.6$\pm$1.6 		& (2.44$\pm$0.29)E$-$16 	& 49.0$\pm$7.3 	& 1890$\pm$170\\
\enddata
\tablecomments{Column (1): Line name, (2): Rest-frame wavelength of the line, (3): Line redshift, (4): Rest-frame FWHM, (5): Line flux, (6): Rest-frame EW, (7): Velocity width after the correction for the instrumental broadening.} 
\end{deluxetable*}
%
%
%J1202
\begin{deluxetable*}{crrrrrr}
\tablecaption{The detected lines of J1202\label{tab:dlineJ1202}}
\tablehead{
\colhead{Line name} & \colhead{$\lambda_{\rm rest}$ [\AA]} & \colhead{$z_{\rm line}$} & \colhead{$FWHM_{\rm rest}$ [\AA]} & \colhead{$F_{\rm line}$ [${\rm erg\ s^{-1}\ cm^{-2}}$]}  & \colhead{$EW_{\rm rest}$ [\AA]} & \colhead{$v_{\rm width}$ [${\rm km\ s^{-1}}$]} \\
\colhead{(1)} & \colhead{(2)} & \colhead{(3)} & \colhead{(4)} & \colhead{(5)} & \colhead{(6)} & \colhead{(7)} 
}
\startdata
Si~{\sc iv} 					& 1393.8 & 2.831$\pm$0.001 	& 4.27$\pm$0.73 	& (1.56$\pm$0.33)E$-$16 	& 5.55$\pm$1.17 	& 919$\pm$157\\
Broad component$^{\rm a}$ 		& ---	       & ---			 	& ---				& (1.23$\pm$0.13)E$-$15 	& --- 				& 4970$\pm$170\\
O~{\sc iv}{]} 					& 1399.9 & 2.842$\pm$0.001 	& 9.90$\pm$0.87 	& (5.13$\pm$0.68)E$-$16 	& 18.1$\pm$2.4 	& 2120$\pm$190\\\hline
C~{\sc iv}\_R 					& 1549.5 & 2.831$\pm$0.001 	& 15.0$\pm$0.4 	& (5.46$\pm$0.26)E$-$15 	& 177$\pm$9 		& 2900$\pm$70\\
C~{\sc iv}\_B 					& 1549.5 & 2.793$\pm$0.002 	& 13.0$\pm$1.0 	& (7.98$\pm$1.03)E$-$16 	& 27.7$\pm$3.6 	& 2510$\pm$190\\
C~{\sc iv}\_R + C~{\sc iv}\_B 		& 1549.5 & 2.830$\pm$0.002 	& 16.0$\pm$0.5 	& (6.26$\pm$0.28)E$-$15 	& 203$\pm$10 		& 3100$\pm$90\\\hline
He~{\sc ii}\_R 					& 1640.4 & 2.838$\pm$0.002 	& 11.0$\pm$1.7 	& (1.68$\pm$0.47)E$-$16 	& 5.06$\pm$1.43 	& 2010$\pm$310\\
He~{\sc ii}\_B 					& 1640.4 & 2.806$\pm$0.005 	& 21.1$\pm$3.4 	& (3.16$\pm$0.55)E$-$16 	& 9.31$\pm$1.64 	& 3870$\pm$620\\
He~{\sc ii}\_R + He~{\sc ii}\_B 		& 1640.4 & 2.834$\pm$0.004 	& 25.5$\pm$2.5 	& (4.84$\pm$0.73)E$-$16 	& 14.5$\pm$2.2 	& 4650$\pm$450\\
O~{\sc iii}{]} 					& 1663.5 & 2.852$\pm$0.003 	& 7.57$\pm$2.94 	& (2.06$\pm$1.07)E$-$17 	& 0.661$\pm$0.343 	& 1360$\pm$530\\\hline
Al~{\sc iii} 						& 1858.8 & 2.839$\pm$0.002 	& 21.5$\pm$2.4 	& (3.05$\pm$0.41)E$-$16 	& 10.5$\pm$1.4 	& 3470$\pm$390\\
Si~{\sc iii} 						& 1892.0 & 2.819$\pm$0.004 	& 13.6$\pm$5.5 	& (9.40$\pm$4.81)E$-$17 	& 3.19$\pm$1.64 	& 2160$\pm$870\\
C~{\sc iii}{]} 					& 1908.7 & 2.835$\pm$0.002 	& 30.6$\pm$1.9 	& (1.01$\pm$0.07)E$-$15 	& 33.5$\pm$2.4	& 4810$\pm$300\\
\enddata
\tablecomments{See Table \ref{tab:dlineJ0907} for the description of each column.   \\{$^{\rm a}$The observed-frame wavelength, continuum flux, and FWHM of the broad component are 5349.5$\pm$2.0 \AA, (7.36$\pm$0.12)E$-$18 ${\rm erg\ s^{-1}\ cm^{-2}}$ \AA$^{-1}$, and 88.9$\pm$2.9 \AA, respectively. See Section \ref{subsec:del} for details.}}
\end{deluxetable*}
%
%
%
%J1207
\begin{deluxetable*}{crrrrrr}
\tablecaption{The detected lines of J1207\label{tab:dlineJ1207}}
\tablehead{
\colhead{Line name} & \colhead{$\lambda_{\rm rest}$ [\AA]} & \colhead{$z_{\rm line}$} & \colhead{$FWHM_{\rm rest}$ [\AA]} & \colhead{$F_{\rm line}$ [${\rm erg\ s^{-1}\ cm^{-2}}$]}  & \colhead{$EW_{\rm rest}$ [\AA]} & \colhead{$v_{\rm width}$ [${\rm km\ s^{-1}}$]} \\
\colhead{(1)} & \colhead{(2)} & \colhead{(3)} & \colhead{(4)} & \colhead{(5)} & \colhead{(6)} & \colhead{(7)} 
}
\startdata
C~{\sc iv}\_R 					& 1549.5 & 2.512$\pm$0.001 & 7.87$\pm$0.43 	& (1.02$\pm$0.08)E$-$15 	& 74.2$\pm$6.0 	& 1520$\pm$80\\
C~{\sc iv}\_B 					& 1549.5 & 2.500$\pm$0.001 & 20.2$\pm$0.5 		& (1.36$\pm$0.09)E$-$15 	& 99.3$\pm$6.9 	& 3900$\pm$90\\
C~{\sc iv}\_R + C~{\sc iv}\_B 		& 1549.5 & 2.511$\pm$0.001 & 10.3$\pm$0.4 		& (2.39$\pm$0.12)E$-$15 	& 173$\pm$9 		& 1990$\pm$70\\\hline
He~{\sc ii}\_R 					& 1640.4 & 2.511$\pm$0.001 & 6.53$\pm$1.12 	& (6.61$\pm$1.50)E$-$17 	& 5.32$\pm$1.21 	& 1190$\pm$210\\
He~{\sc ii}\_B 					& 1640.4 & 2.499$\pm$0.002 & 17.7$\pm$1.3 		& (1.47$\pm$0.24)E$-$16 	& 11.7$\pm$1.9 	& 3240$\pm$240\\
He~{\sc ii}\_R + He~{\sc ii}\_B 		& 1640.4 & 2.509$\pm$0.002 & 11.6$\pm$1.2 		& (2.13$\pm$0.28)E$-$16 	& 17.2$\pm$2.3 	& 2120$\pm$220\\
O~{\sc iii}{]} 					& 1663.5 & 2.516$\pm$0.002 & 12.2$\pm$2.4 		& (3.70$\pm$0.95)E$-$17 	& 3.18$\pm$0.82 	& 2200$\pm$430\\\hline
Al~{\sc iii} 						& 1858.8 & 2.508$\pm$0.002 & 19.7$\pm$1.7 		& (9.09$\pm$1.02)E$-$17 	& 9.23$\pm$1.04 	& 3180$\pm$280\\
Si~{\sc iii} 						& 1892.0 & 2.511$\pm$0.003 & 10.7$\pm$4.9 		& (2.64$\pm$1.60)E$-$17 	& 2.79$\pm$1.69 	& 1690$\pm$770\\
C~{\sc iii}{]}\_R 					& 1908.7 & 2.510$\pm$0.001 & 12.9$\pm$1.4 		& (1.69$\pm$0.25)E$-$16 	& 18.2$\pm$2.7 	& 2030$\pm$220\\
C~{\sc iii}{]}\_B 					& 1908.7 & 2.503$\pm$0.002 & 42.0$\pm$2.9 		& (4.02$\pm$0.65)E$-$16 	& 43.1$\pm$7.0 	& 6600$\pm$450\\
C~{\sc iii}{]}\_R + C~{\sc iii}{]}\_B 	& 1908.7 & 2.509$\pm$0.002 & 19.6$\pm$1.2 		& (5.71$\pm$0.70)E$-$16 	& 61.4$\pm$7.5 	& 3080$\pm$180\\
\enddata
\tablecomments{See Table \ref{tab:dlineJ0907} for the description of each column.}
\end{deluxetable*}
%
%
%J1207
\begin{deluxetable*}{crrrrrr}
\tablecaption{The detected lines of J1443\label{tab:dlineJ1443}}
\tablehead{
\colhead{Line name} & \colhead{$\lambda_{\rm rest}$ [\AA]} & \colhead{$z_{\rm line}$} & \colhead{$FWHM_{\rm rest}$ [\AA]} & \colhead{$F_{\rm line}$ [${\rm erg\ s^{-1}\ cm^{-2}}$]}  & \colhead{$EW_{\rm rest}$ [\AA]} & \colhead{$v_{\rm width}$ [${\rm km\ s^{-1}}$]} \\
\colhead{(1)} & \colhead{(2)} & \colhead{(3)} & \colhead{(4)} & \colhead{(5)} & \colhead{(6)} & \colhead{(7)} 
}
\startdata
Ly~${\alpha}$ 					& 1215.7 & 3.341$\pm$0.003 & 10.7$\pm$2.6 		& (5.28$\pm$1.31)E$-$16 	& 64.3$\pm$17.3 	& 2650$\pm$640\\
N~{\sc v} 						& 1240.8 & 3.312$\pm$0.001 & 16.8$\pm$0.6 		& (2.06$\pm$0.09)E$-$15 	& 242$\pm$23 		& 4070$\pm$160\\
Si~{\sc ii} 						& 1262.6 & 3.326$\pm$0.003 & 16.0$\pm$1.9 		& (1.24$\pm$0.17)E$-$16 	& 13.6$\pm$2.0 	& 3790$\pm$450\\\hline
Si~{\sc iv} 					& 1393.8 & 3.324$\pm$0.010 & 16.4$\pm$3.2 		& (3.18$\pm$1.01)E$-$16 	& 33.8$\pm$10.7 	& 3520$\pm$700\\
O~{\sc iv}{]} 					& 1399.9 & 3.339$\pm$0.008 & 12.9$\pm$2.5 		& (1.74$\pm$1.11)E$-$16 	& 18.6$\pm$11.8 	& 2760$\pm$540\\\hline
C~{\sc iv}\_R 					& 1549.5 & 3.328$\pm$0.005 & 14.2$\pm$1.6 		& (6.15$\pm$1.74)E$-$16 	& 60.0$\pm$17.0 	& 2740$\pm$310\\
C~{\sc iv}\_B 					& 1549.5 & 3.296$\pm$0.007 & 15.7$\pm$2.3 		& (5.47$\pm$1.67)E$-$16 	& 54.9$\pm$16.8 	& 3030$\pm$440\\
C~{\sc iv}\_R + C~{\sc iv}\_B 		& 1549.5 & 3.317$\pm$0.006 & 23.1$\pm$1.8 		& (1.16$\pm$0.24)E$-$15 	& 114$\pm$24 		& 4470$\pm$350\\\hline
He~{\sc ii}\_R 					& 1640.4 & 3.337$\pm$0.001 & 6.48$\pm$1.56 	& (3.23$\pm$1.01)E$-$17 	& 3.24$\pm$1.02 	& 1180$\pm$290\\
He~{\sc ii}\_B 					& 1640.4 & 3.307$\pm$0.004 & 20.3$\pm$2.9 		& (9.15$\pm$1.53)E$-$17 	& 9.25$\pm$1.56 	& 3710$\pm$530\\
He~{\sc ii}\_R + He~{\sc ii}\_B 		& 1640.4 & 3.335$\pm$0.003 & 20.1$\pm$2.5 		& (1.24$\pm$0.18)E$-$16 	& 12.4$\pm$1.9 	& 3680$\pm$460\\
\enddata
\tablecomments{See Table \ref{tab:dlineJ0907} for the description of each column.}
\end{deluxetable*}
%%%%%%%%%%%%%%%%%%%%%%%%%%%%%%%

\begin{deluxetable*}{ccccccc}
\tablecaption{Emission-line contribution and excess magnitude to the power-law fit in the $g$- and $r$-band \label{tab:EVEWS}}
\tablehead{
\colhead{} & \multicolumn{3}{c}{$g$-band} & \multicolumn{3}{c}{$r$-band}\\
\colhead{} & \colhead{Total EWs} & \colhead{$\Delta$mag} & \colhead{Excess mag} & \colhead{Total EWs} & \colhead{$\Delta$mag} & \colhead{Excess mag}\\
\colhead{} & \colhead{[\AA]} & \colhead{[AB mag]} & \colhead{[AB mag]} & \colhead{[\AA]} & \colhead{[AB mag]} & \colhead{[AB mag]}
}
\startdata
HSC J0907 	& $>$604 	& $>$0.37 	& 1.27 	& 129 	& 0.09 	& 0.47\\
HSC J1202	& $>$258 	& $>$0.18	 	& 1.13	& 836	& 0.48	& 0.44\\
HSC J1207	& $>$608	& $>$0.38		& 1.48	& 329	& 0.21	& 0.65\\
HSC J1443	& 1380	& 0.72		& 0.88	& 721	& 0.43	& 0.40\\\hline
\enddata
\tablecomments{$\Delta$mag: $2.5\log{(1+EW/BW)}$, Excess mag: The excesses of the $g$- and $r$- band magnitudes between the observed magnitudes and the expected magnitudes from an extrapolation of the power-law fit to the longer wavelength bands.}
\end{deluxetable*}

\begin{figure}%[h]
	%\centering
		\includegraphics[width=8.5cm]{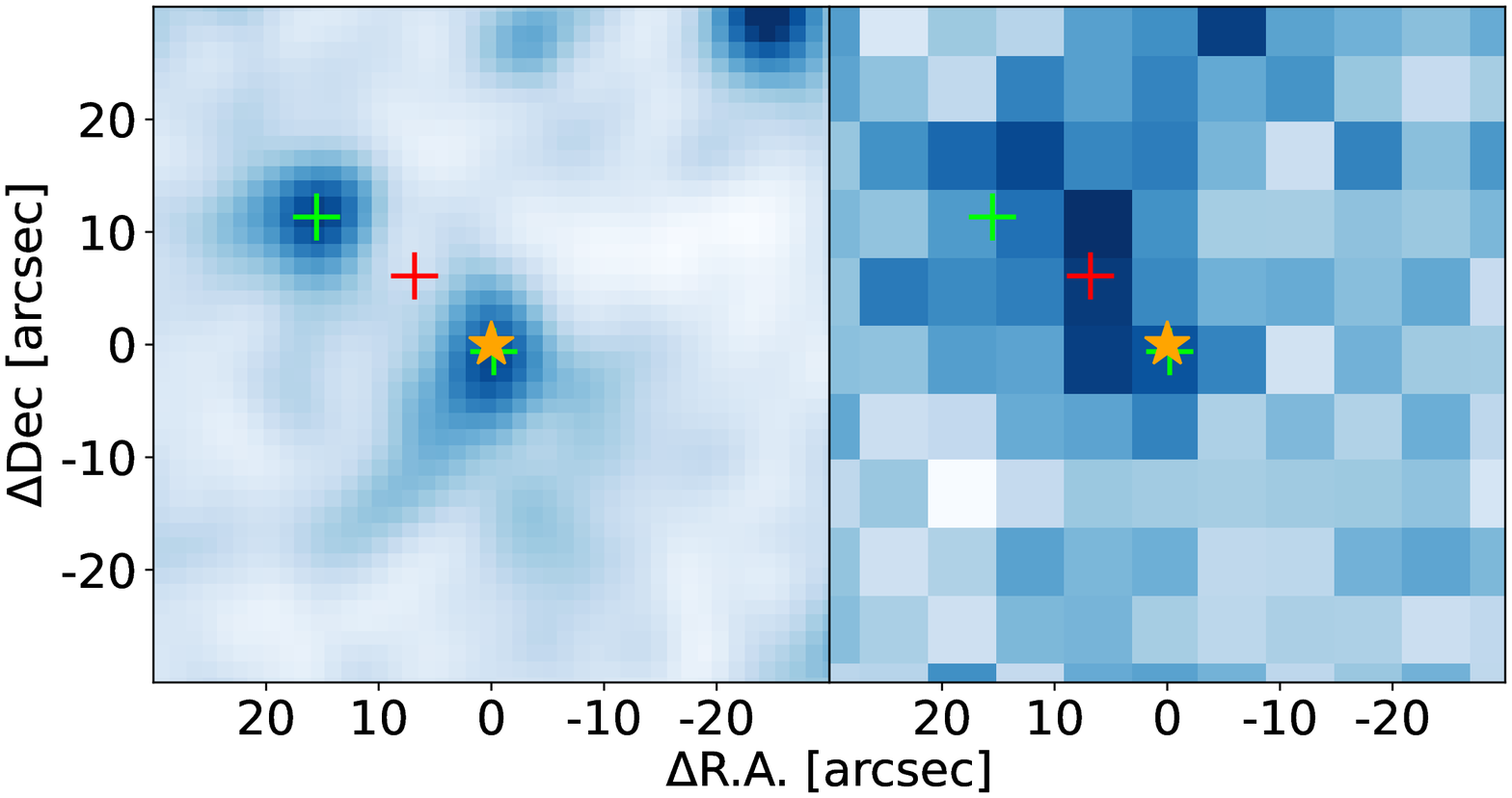} 
		%\vspace{30pt}
		\caption{J1207 images in the {\it WISE} $W1$-band (left) and H-ATLAS 250 $\mu$m-band (right). The orange stars, green crosses, and red cross denote source detections in the HSC-SSP, ALLWISE, and H-ATLAS catalogs, respectively. The size of each image is 60\arcsec $\times$ 60\arcsec, centered at the HSC position of J1207.}
		\label{fig:J1207cp}
\end{figure}

\subsection{Estimating the dust extinction}\label{subsec:CIGALECODE}
We need to estimate dust extinction, $E(B-V)$ of AGN radiation, and $L^{\rm AGN}_{\rm bol}$ to calculate the SMBH mass and Eddington ratio.
Since Balmer decrement or other spectral measures of $E(B-V)$ is not available,
we perform the SED fitting to the broad-band photometry to estimate the $E(B-V)$ and $L^{\rm AGN}_{\rm bol}$.
In this work, we utilize the new version of {\tt Code Investigating GAlaxy Emission} ({\tt CIGALE}; \citealt{2005MNRAS.360.1413B, 2009A&A...507.1793N, 2019A&A...622A.103B}) called {\tt X-CIGALE} \citep{2020MNRAS.491..740Y}, to perform the SED fit in a self-consistent framework by considering an energy balance between the UV/optical absorption and IR emission. 
{\tt X-CIGALE} generates the best-fit model including the stellar, AGN, and SF components that fits the photometric data in the rest-frame UV to far-infrared (FIR) bands.
We utilize the {\it Herschel} Space Observatory (\citealt{2010A&A...518L...1P}) Astrophysical Terahertz Large Area Survey (H-ATLAS; \citealt{2010PASP..122..499E, 2016MNRAS.462.3146V, 2016MNRAS.462.1714B}) data observed with Photodetector Array Camera and Spectrometer (PACS; \citealt{2010A&A...518L...2P}) at 100 and 160 $\mu$m and with Spectral and Photometric Imaging REceiver (SPIRE; \citealt{2010A&A...518L...3G}) at 250, 350, and 500 $\mu$m in the FIR, in addition to optical, NIR, and MIR data obtained by Subaru HSC, VISTA, and WISE. 
The 1$\sigma$ limiting fluxes at 100, 160, 250, 350, and 500 $\mu$m are 44, 49, 7.4, 9.4, and 10.2 mJy, respectively (\citealt{2016MNRAS.462.3146V}).

To search for the H-ATLAS counterpart of the four BluDOGs, we adopt a search radius of 10 arcsec by following \cite{2019ApJS..243...15T} (and \citealt{2022A&A...661A..15T}).
Accordingly we found the counterparts of two BluDOGs (J1202 and J1207).
The separation between the HSC position and the H-ATLAS counterpart position is 0.94 arcsec for J1202, and 9.7 arcsec for J1207.
The relatively large separation in the latter case suggests the counterpart being a coincidental detection.
There are two {\it WISE} sources around J1207 (Figure~\ref{fig:J1207cp}); one probably corresponds to J1207 itself (the angular separation between the HSC and {\it WISE} potions is 0.65 arcsec) and another is located at 19 arcsec away to the north-east direction. 
The H-ATLAS source is located between these two {\it WISE} sources, and thus the FIR fluxes given in the H-ATLAS catalog are possibly attributed to the two {\it WISE} sources. 
Therefore we regard the H-ATLAS fluxes of J1207 as the upper limit. 
For the remaining two BluDOGs (J0907 and J1443), we adopt the  5$\sigma$ upper limit fluxes.

As for the optical--MIR photometric data, we utilize $g$, $r$, $i$, $z$, $y$ (HSC-SSP), $Z$, $Y$, $J$, $H$, $Ks$ (VIKING DR2), $W1$, $W2$, $W3$, and $W4$ (ALLWISE) bands (see \citealt{2019ApJ...876..132N}).
Note that the SNRs in these bands are more than 5, except for the $W4$-band with SNR more than 3 because we adopted such SNR cut in the selection of DOGs (\citealt{2019ApJ...876..132N}).
Since the $g$- and $r$-band photometry are significantly affected by the strong emission lines (see Figure \ref{fig:bludogsp} and Tables \ref{tab:dlineJ0907}--\ref{tab:dlineJ1443}) which cannot be treated properly in {\tt X-CIGALE}, we corrected for their contribution by referring to the estimates given in Table~\ref{tab:EVEWS}.

\begin{deluxetable}{cc}
\tablecaption{Parameters adopted in the {\tt X-CIGALE} fit \label{tab:cigale}}
\tablehead{
\colhead{Parameter}	& \colhead{Value} 
}
\startdata
\multicolumn{2}{c}{Delayed SFH (\citealt{2015AandA...576A..10C})} \\\hline
${\tau}_{\rm main}$ [Myr] & 100, 250, 500\\
${\tau}_{\rm burst}$ [Myr] & 10, 50\\
$f_{\rm burst}$ & 0.0, 0.5, 0.99\\
Age$_{\rm main}$ [Myr] & 500, 800, 1000\\
Age$_{\rm burst}$ [Myr] & 1, 5, 10\\\hline
\multicolumn{2}{c}{Single stellar population (\citealt{2003MNRAS.344.1000B})} \\\hline
IMF & \cite{2003PASP..115..763C}\\
Metallicity & 0.02\\
Age$_{\rm separation}$ [Myr]& 10\\\hline
\multicolumn{2}{c}{Nebular emission (\citealt{2011MNRAS.415.2920I})} \\\hline
$\log{U}$ & $-$2.0\\
$f_{\rm esc}$ & 0.0\\
$f_{\rm dust}$ & 0.0\\
Lines width [${\rm km\ s^{-1}}$]& 300.0\\\hline
\multicolumn{2}{c}{Dust attenuation (\citealt{2000ApJ...533..682C})} \\\hline
$E(B-V)_{\rm line}$ & 3, 4, 5, 6, 7, 8, 9, 10 \\
$f_{E(B-V)}$ & 0.44\\
$\lambda_{\rm UV, bump}$ [nm] & 217.5\\
FWHM$_{\rm UV, bump}$ [nm] & 35.0\\
$A_{\rm UV, bump}$ & 0.0\\
$\delta$ & 0.0\\
Extinction law of emission lines & the Milky Way\\
$R_V$ & 3.1\\\hline
\multicolumn{2}{c}{Dust emission (\citealt{2014ApJ...784...83D})} \\\hline
AGN fraction & 0.0\\
$\alpha_{\rm IR, AGN}$ & 0.0625, 0.2500, 2.0000\\\hline
\multicolumn{2}{c}{AGN model (\citealt{2016MNRAS.458.2288S})} \\\hline
$\tau_{9.7}$ & 3, 7\\
$p$ & 1.0\\
$q$ & 1.0\\
$oa$ [deg] & 10, 20, 30, 40, 50, 60, 70, 80 \\
$R_{\rm ratio}$ & 20\\
$M_{\rm cl}$ & 0.97\\
$i$ [deg] & 0, 10, 20, 30, 40,\\
& 50, 60, 70, 80, 90\\
$f_{\rm AGN}$ & 0.1, 0.3, 0.5, 0.7, 0.9\\
Extinction law of polar dust & \citealt{2000ApJ...533..682C}\\
$E(B-V)^{\rm AGN}_{\rm polar\ dust}$ & 0.1, 0.2, 0.3, 0.4, 0.5 \\
$T^{\rm AGN}_{\rm polar\ dust}$ [K] & 600, 700, 800, 900,\\
& 1000, 1100, 1200, 1300, 1400\\
Emissivity of polar dust & 1.6\\\hline
\enddata
\end{deluxetable}

The models and parameters of {\tt X-CIGALE} adopted in this work are summarized in Table \ref{tab:cigale}.
We assume a delayed star formation history (SFH; \citealt{2015AandA...576A..10C}) with the e-folding times of the main stellar population ($\tau_{\rm main}$) and late starburst population ($\tau_{\rm burst}$), mass fraction of the late burst population ($f_{\rm burst}$), and age of the main stellar population (Age$_{\rm main}$) and the late burst (Age$_{\rm burst}$).
As the stellar population, we assume the initial mass function of \cite{2003PASP..115..763C}, solar metallicity, and 10-Gyr separation between young and old stellar population (Age$_{\rm separation}$).
The nebular emission model (\citealt{2011MNRAS.415.2920I}) is characterized by the ionization parameter ($U$), fractions of Lyman continuum photons escaping the galaxy ($f_{\rm esc}$) and absorbed by dust ($f_{\rm dust}$), and line width.
We utilize a modified dust attenuation model presented by \cite{2019A&A...622A.103B}.
The dust attenuation model for the continuum is taken from \cite{2000ApJ...533..682C} with the extension taken from \cite{2002ApJS..140..303L} between the Lyman break and 1500 {\AA}.
The emission lines are attenuated with a Milky Way extinction with $R_V = 3.1$ \citep{1989ApJ...345..245C}.
We assumed $E(B-V)_{\rm continuum} = 0.44 E(B-V)_{\rm line}$, following \cite{2000ApJ...533..682C}.
The $E(B-V)_{\rm line}$ is varied between 3 and 10.
We utilize the {\tt SKIRTOR} model as the AGN emission model, which takes geometric parameters of the AGN into account and also allows us to incorporate the effect of extinction by the polar dust.
The parameters of the AGN model are the average edge-on optical depth at 9.7 $\mu$m ($\tau_{9.7}$), the torus density parameters ($p$ and $q$; \citealt{2016MNRAS.458.2288S}), the angle between the equatorial plane and the edge of the torus ($oa$), the ratio of the maximum to minimum radii of the dust torus ($R_{\rm ratio}$), the fraction of total dust mass inside clumps ($M_{\rm cl}$), the inclination ($i$), the AGN fraction ($f_{\rm AGN}$), the extinction law, color excess ($E(B-V)^{\rm AGN}_{\rm polar\ dust}$), dust temperature ($T^{\rm AGN}_{\rm polar\ dust}$), and emissivity index of the polar dust.

\begin{figure*}%[h]
	%\centering
	%\vspace{-8cm}
	\includegraphics[width=18cm]{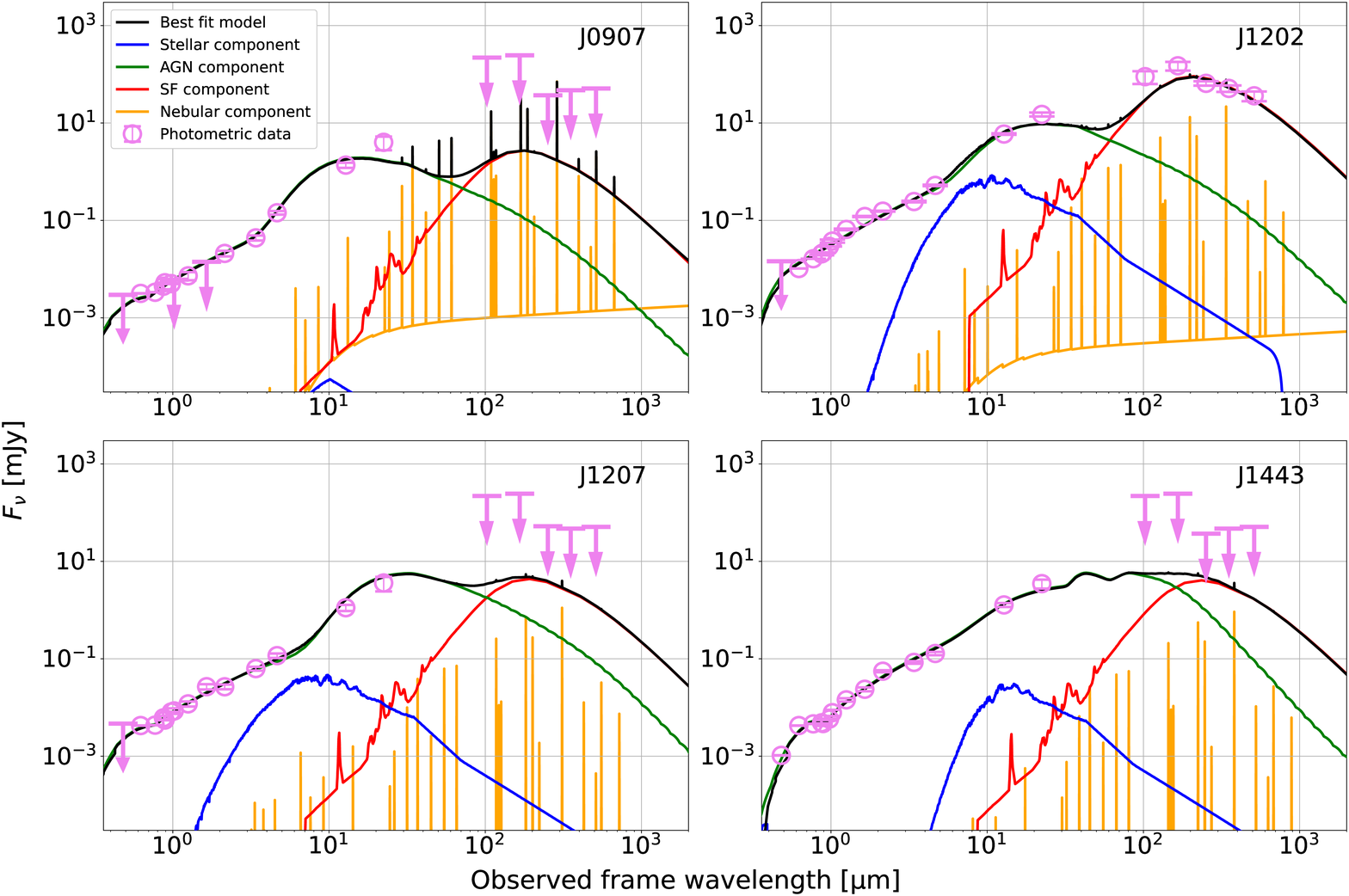} 
		%\vspace{30pt}
	\caption{The results of the SED fitting for the four BluDOGs. The upper left, upper right, lower left and lower right panels show the results of J0907, J1202, J1207, and J1443, respectively. The black, blue, green, red, and orange lines represent the best-fit model, stellar component (with dust attenuation), AGN component, SF component (FIR re-emission from the dust heated by SF), and nebular component, respectively. The magenta plots represent the photometric data. The arrows denote 5$\sigma$ upper limit flux.}
	\label{fig:CIGALERESULT}
\end{figure*}

The best-fit SED models are shown in Figure \ref{fig:CIGALERESULT}.
The reduced $\chi^2$ of the fits are 1.38, 3.19, 0.93, and 1.74 for J0907, J1202, J1207, and J1443, respectively.
The best-fit values and associated errors for $E(B-V)^{\rm AGN}_{\rm polar\ dust}$ and $L^{\rm AGN}_{\rm bol}$ are estimated with a Bayesian-like strategy presented in \cite{2009A&A...507.1793N}, and are reported in Table~\ref{tab:BHP}.
On the other hands, we cannot quantitatively constrain the parameters of the host galaxies because the $E(B-V)$ values are too large and the optical parts in their SEDs are dominated by their AGN emission (see Figure~\ref{fig:CIGALERESULT}).

\subsection{Measurement of the SMBH mass}\label{subsec:M_SMBH_MASS}

\begin{deluxetable*}{ccccc}
\tablecaption{Physical properties of the four BluDOGs\label{tab:BHP}}
\tablehead{
\colhead{} & \colhead{HSC J0907} & \colhead{HSC J1202} & \colhead{HSC J1207} & \colhead{HSC J1443}
}
\startdata
Redshift							& $2.258\pm0.002$				& $2.830\pm0.002$				& $2.511\pm0.001$				& $3.317\pm0.006$				\\
$E(B-V)^{\rm AGN}_{\rm polar\ dust}$	& $0.26\pm0.05$				& $0.33\pm0.04$				& $0.30\pm0.02$				& $0.20\pm0.01$				\\
$L^{\rm AGN}_{\rm bol}$/$L_\odot$		& $(6.11\pm0.95)\times10^{12}$	& $(6.11\pm1.18)\times10^{13}$	& $(7.95\pm1.47)\times10^{12}$	& $(2.52\pm0.13)\times10^{13}$	\\
$M_{\rm BH}$ (C~{\sc iv})/$M_\odot$	& $(1.69\pm0.17)\times10^{8}$		& $(4.95\pm0.30)\times10^{8}$		& $(1.11\pm0.08)\times10^{8}$		& $(5.48\pm0.86)\times10^{8}$		\\
$M_{\rm BH}$ (Mg~{\sc ii})/$M_\odot$	& $(9.85\pm1.80)\times10^{7}$		& ---							& ---							& ---							\\
$\lambda_{\rm Edd}$ (C~{\sc iv}) 		& $1.10\pm0.20$				& $3.75\pm0.76$				& $2.19\pm0.43$				& $1.40\pm0.23$				\\\hline
\enddata
\end{deluxetable*}

We have detected the C~{\sc iv} emission line for all the four BluDOGs and Mg~{\sc ii} emission line for J0907, both of which are widely used to calculate the SMBH mass of type-1 AGNs.
Note that the systematic uncertainty is larger in the C~{\sc iv}-based SMBH mass than in the Mg~{\sc ii}-based SMBH mass, due to a powerful outflow sometimes seen in the C~{\sc iv} velocity profile \citep[e.g.,][]{2005MNRAS.356.1029B, 2015ARA&A..53..365N, 2017MNRAS.465.2120C}.
We calculate the single-epoch mass of SMBHs with the C~{\sc iv} and Mg~{\sc ii} emission lines, following the calibrations given in \cite{2006ApJ...641..689V} and \cite{2009ApJ...699..800V} respectively:
\begin{eqnarray}
   M_{\rm BH} &=& 10^{6.66}\Big(\frac{{\rm FWHM(C~{\sc IV})}}{10^3\ {\rm km\ s^{-1}}}\Big)^2\Big(\frac{\lambda L_\lambda(1350{\rm \AA})}{10^{44}\ {\rm erg\ s^{-1}}}\Big)^{0.53}M_\odot,\nonumber\\
   &&
\end{eqnarray}
and
\begin{eqnarray}
   M_{\rm BH} &=& 10^{6.86}\Big(\frac{{\rm FWHM(Mg~{\sc II})}}{10^3\ {\rm km\ s^{-1}}}\Big)^2\Big(\frac{\lambda L_\lambda(3000{\rm \AA})}{10^{44}\ {\rm erg\ s^{-1}}}\Big)^{0.5}M_\odot,\nonumber\\
   &&
\end{eqnarray}
where FWHM(C~{\sc iv}), FWHM(Mg~{\sc ii}), $\lambda L_\lambda(1350{\rm \AA})$ and $\lambda L_\lambda(3000{\rm \AA})$ are the FWHM of the C~{\sc iv} and Mg~{\sc ii} velocity profile, and the monochromatic luminosity at 1350 {\AA} and 3000 {\AA}, respectively.
Note that we use the FWHM of C~{\sc iv}\_R + C~{\sc iv}\_B as the FWHM of the C~{\sc iv}.
We cannot eliminate the possiblility that the estimated SMBH masses are overestimated because the C~{\sc iv} profiles are affected by nucleus outflows (Section~\ref{subsec:SpecFeature}).
For estimating the reddening-corrected monochromatic luminosity, we use the optical spectra presented in Section \ref{subsec:del}. 
We converted the spectra to the rest-frame, de-reddened them with $E(B-V)^{\rm AGN}_{\rm polar\ dust}$ derived in the SED fit, and masked out emission and absorption lines as well as pixels with negative values.
Then, we fit a power-law continuum model to the spectra and estimate the monochromatic luminosities from the best fits.
The estimated $\lambda L_{\lambda}$(1350) of J0907, J1202, J1207, and J1443 are $(1.54\pm0.05)\times10^{45}$, $(9.64\pm0.27)\times10^{45}$, $(3.06\pm0.06)\times10^{45}$, and $(2.93\pm0.03)\times10^{45}$ ${\rm erg\ s^{-1}}$, respectively.
The $\lambda L_{\lambda}$(3000) of J0907 is estimated to be $(1.45\pm0.04)\times10^{45}$ ${\rm erg\ s^{-1}}$.

The resultant SMBH masses are summarized in Table~\ref{tab:BHP}.
It should be noted that the C~{\sc iv}-based $M_{\rm BH}$ and Mg~{\sc ii}-based $M_{\rm BH}$ of J0907 is not consistent within the statistical error.
This is probably attributed to a systematic error especially in the C~{\sc iv}-based $M_{\rm BH}$, known to be accompanied with a large systematic error ($\sim$0.5 dex; see, e.g., \citealt{2013BASI...41...61S}). 
Hereafter we use only the C~{\sc iv}-based $M_{\rm BH}$, since it is measured in all the four BluDOGs.

%%%%%%%%%%%%%%%%%%%%%%%%%%%%
\section{Discussion} \label{sec:disscussion}

\subsection{Spectral features and nuclear outflows}\label{subsec:SpecFeature}

We found that the redshifts of the four BluDOGs are in the range of $2.2 \lesssim z_{\rm sp} \lesssim 3.3$. 
They are systematically higher than the typical redshifts of DOGs ($z_{\rm sp} = 1.99\pm0.45$; \citealt{2008ApJ...677..943D, 2008ApJ...689..127P}). 
One possible reason for this systematically high redshift is a selection effect related to the blue-excess criterion. 
When we select BluDOGs from the parent DOG sample, the $g$- and $r$-band magnitudes show an excess of the expected magnitudes estimated by the power-law extrapolation from $i$-band to $W4$-band.
Thus we may select DOGs in a preferred redshift range where strong emission lines such as Ly$\alpha$ and C~{\sc iv} shifts into the two bands (see Section~\ref{subsec:CIVEW} for more quantitative assessments).
The reason for the underestimated photometric redshift ($\sim 1$;  \citealt{2019ApJ...876..132N}) is the unusual emission lines with the large REW.

The detected emission lines have large velocity widths, $\gtrsim 2000\ {\rm km\ s^{-1}}$ in most cases. 
This suggests that the broad-line region (BLR) of the BluDOGs is not completely obscured; in other words, the observed BluDOGs are classified as type-1 AGNs. 
This is an unexpected result, because their very red color between optical and mid-IR suggests the heavily obscured nature. 
One possible interpretation is that we are looking at a phase where the surrounding dust is just blown away by the nuclear activity (outflow, radiation pressure, or both), as discussed more in Section~\ref{subsec:DBHER}. 
It should be noted that the type-1 nature is seen not only in the presented BluDOGs but also in some other DOGs \citep[e.g.,][]{2016ApJ...820...46T, 2017ApJ...850..140T, 2020MNRAS.499.1823Z}. 
Systematic spectroscopic observations for the whole populations of DOGs are required to study the nature of obscuration occurring in various populations of DOGs.

As shown in Figure~\ref{fig:BluCIV}, the velocity profile of the observed C~{\sc iv} lines show a notable excess feature in the blue wing.
Such an excess in the C~{\sc iv} velocity profile has been observed in other type-1 AGNs, and interpreted as a result of powerful nuclear outflows \citep[e.g.,][]{2005MNRAS.356.1029B, 2015ARA&A..53..365N, 2017MNRAS.465.2120C}. To evaluate quantitatively how the nuclear outflow in BluDOGs is strong compared to ordinary AGNs, we examine the ``asymmetry parameter ($\alpha_\beta$)'' defined by \cite{1985ApJ...289...67D} as 
\begin{eqnarray}
   \alpha_\beta &=& \frac{\lambda_c(3/4) - \lambda_c(1/4)}{\Delta\lambda(1/2)},
\end{eqnarray}
where $\lambda_c(h)$ and $\Delta\lambda(1/2)$ are the central wavelength at which the flux falls to a $h$ time the peak flux and FWHM of the broad profile, respectively. 
The positive and negative values of $\alpha_\beta$ express the blue and red excesses, respectively.
The derived values of $\alpha_\beta$ for J0907, J1202, J1207, and J1443 are 0.216, 0.102, 0.246, and 0.051, respectively. As a reference, the C~{\sc iv} velocity profile in the composite spectrum of SDSS type-1 quasars given by \cite{2001AJ....122..549V} shows $\alpha_\beta = 0.110$. 
Thus J0907, and J1207 may possess a significant nuclear outflow that is more powerful than typical quasars.

\begin{figure}[ht!]
	%\centering
	        \hspace{-0.5cm}
		\includegraphics[width=9.0cm]{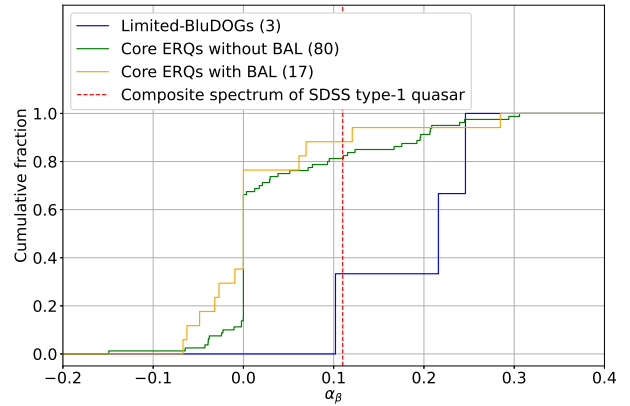} 
		%\vspace{30pt}
		\caption{Cumulative distribution of the $\alpha_\beta$ indices for the limited-BluDOGs (see the main text; blue line), core ERQs without BAL (green line), and core ERQs with BAL (orange line). The red dashed line denotes the $\alpha_\beta$ index measured for the composite spectrum of SDSS type-1 quasars.}
		\label{fig:ABC}
\end{figure}

In order to compare $\alpha_\beta$ of the BluDOG with that of another dusty AGN population, we fitted the C~{\sc iv} profile of 97 ``core'' ERQs (ERQs with REW(C~{\sc iv}) $>100$ \AA) in \cite{2017MNRAS.464.3431H} and measured $\alpha_\beta$ by adopting a single or double Gaussian profile.
The core ERQ sample consists of 80 objects without BAL and 17 objects with BAL, and we investigate the statistics of $\alpha_\beta$ for the two subsamples separately because the BAL feature can affect the C~{\sc iv} line profile.
Here we exclude J1443 from the BluDOG sample when comparing the $\alpha_\beta$ index because its velocity profile is largely affected by narrow absorption lines (hereafter the limited-BluDOG sample to infer the 3 BluDOGs; i.e., J0907, J1202, and J1207). 
Figure~\ref{fig:ABC} shows the cumulative fraction of $\alpha_\beta$ for the limited-BluDOGs, core ERQs without BAL, and core ERQs with BAL. 
The averaged values of the limited-BluDOGs, core ERQs without BAL, and core ERQs with BAL are 0.15$\pm$0.08, 0.02$\pm$0.13, and 0.01$\pm$0.09, respectively. 
We performed the Kolmogorov-Smirnov test (KS-test) to examine the statistical significance of the difference in $\alpha_\beta$ among the samples.
The {\it p}-values of the limited-BluDOGs-core ERQs without BAL, and limited-BluDOG-core ERQs with BAL are 0.0178 and 0.0175, respectively.
Thus we conclude that the distributions of $\alpha_\beta$ of the limited-BluDOGs and core ERQs with/without BAL are marginally different with $>2$ sigma significance.
This suggests that the BluDOGs show nuclear outflow that is possibly more powerful than the nuclear outflow in core ERQs with/without BAL.

\begin{figure}[ht!]
	%\centering
	        \hspace{-0.5cm}
		\includegraphics[width=9.0cm]{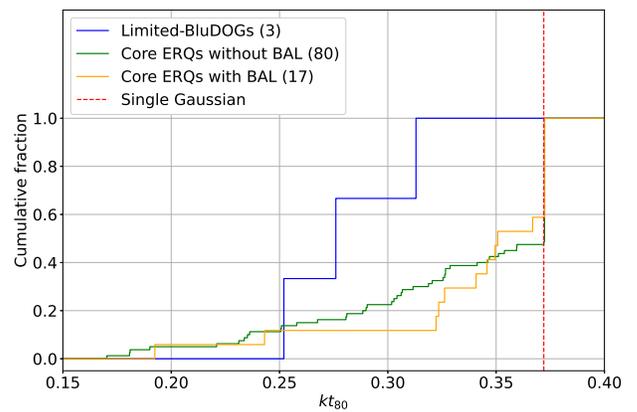} 
		%\vspace{30pt}
		\caption{Cumulative distribution of the $kt_{80}$ indices for the limited-BluDOGs (blue line), core ERQs without BAL (green line), and core ERQs with BAL (orange line). The red dashed line denotes the $kt_{80}$ index for the single Gaussian profile.}
		\label{fig:kt80}
\end{figure}

We also focus on the kurtosis index ($kt_{80}$) defined as follows (see \citealt{2017MNRAS.464.3431H} for detailes): $kt_{80} = \Delta v$(80\%) $/ \Delta v$(20\%), where $\Delta v$($x$\%) is the velocity width at $x$\% of the peak flux height.
In addition to $\alpha_\beta$, this $kt_{80}$ index is useful to characterize the C~{\sc iv} wing (a more prominent blue wing results in smaller $kt_{80}$).
By using the best-fit double Gaussian profile of the BluDOGs, $kt_{80}$ of J0907, J1202, J1207, and J1443 are 0.276, 0.313, 0.252, and 0.440, respectively.
Again we exclude J1443 from the BluDOG sample when comparing the $kt_{80}$ index as the discussion of the $\alpha_\beta$ index.
For comparison, $kt_{80}$ of a single Gaussian is $\sqrt{1-\frac{2\ln(2)}{\ln(5)}}$ ($\sim0.37$), whereas most quasars have $kt_{80}\sim 0.15$--$0.30$ (see Figure~7 in \citealt{2017MNRAS.464.3431H}).
The limited-BluDOGs, core ERQs without BAL, and core ERQs with BAL show $kt_{80} = 0.28\pm0.03$, $0.33\pm0.06$, and $0.34\pm0.05$, respectively (see also Figure~\ref{fig:kt80}).
Note that the C~{\sc iv} profile of 41 core ERQs without BAL and 7 core ERQs with BAL is fitted by a single Gaussian, which is the reason why many objects have $kt_{80}\sim0.37$ as shown in Figure~\ref{fig:kt80}.
C~{\sc iv} velocity profiles of core ERQs with/without BAL are roughly consistent with the Gaussian without a blue wing.
However, the $kt_{80}$ index of the limited-BluDOGs is less than $\sqrt{1-\frac{2\ln(2)}{\ln(5)}}$, suggesting that their C~{\sc iv} line profile has a wing.
We performed the KS-test  to examine the statistical significance of the difference in $kt_{80}$ among the samples.
The {\it p}-values of the limited-BluDOGs-core ERQs without BAL, and limited-BluDOGs-core ERQs with BAL are 0.0637 and 0.0175, respectively.
Therefore, we conclude that the distributions of $kt_{80}$ between the samples of the limited-BluDOGs and core ERQs with/without BAL feature are marginally different with $>2$ sigma significance.

\begin{figure}[ht!]
	%\centering
	        \hspace{-0.5cm}
		\includegraphics[width=9.0cm]{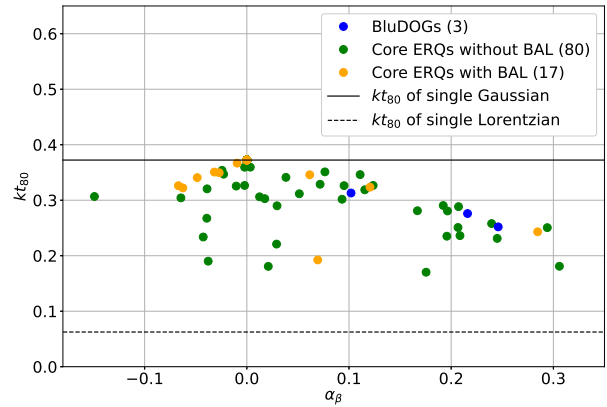} 
		%\vspace{30pt}
		\caption{The $kt_{80}$ vs. $\alpha_\beta$ plot for the limited-BluDOGs (blue dots), core ERQs without BAL (green dots), and core ERQs with BAL (orange dots). The black solid line and dashed line denote the $kt_{80}$ values of single Gaussian and single Lorentzian, respectively. Note that 41 core ERQs without BAL and 7 core ERQs with BAL are fitted with a single Gaussian, and they are plotted at ($\alpha_\beta$, $kt_{80}$) = (0, $\sqrt{1-\frac{2\ln(2)}{\ln(5)}}$).}
		\label{fig:ABC_kt80}
\end{figure}

It has been reported that AGNs with a high Eddington ratio tend to show a Lorentzian-line velocity profile in BLR lines (e.g., \citealt{1996ApJS..106..341M, 2001A&A...372..730V, 2006A&A...456...75C, 2010MNRAS.403.1759Z}).
Therefore, the small $kt_{80}$ value of BluDOGs can be caused by the contribution of extended Lorentzian wings instead of the asymmetric blue wing.
For a symmetric Lorentzian profile, $kt_{80}\sim1/16$ (much smaller than a Gaussian profile, $kt_{80}\sim0.37$) and $\alpha_\beta = 0$ are expected.
However, the BluDOGs are inconsistent with this expectation (Figure~\ref{fig:ABC_kt80}).
This Figure~\ref{fig:ABC_kt80} also shows that the BluDOGs follow the trend made by core ERQs with/without BAL in the $kt_{80} - \alpha_\beta$ plane, while a systematic deviation of BluDOGs toward ($\alpha_\beta$, $kt_{80}$) = (0, 0) is expected if a Lorentzian component significantly contributes to the C~{\sc iv} line of BluDOGs.
Thus, we conclude that extended Lorentzian wings do not affect the C~{\sc iv} line profile of BluDOGs, but the small $kt_{80}$ of BluDOGs is caused by the asymmetric blue excess due to the stronger nuclear outflow than that of ERQs.

\subsection{Large equivalent widths of the CIV emission}\label{subsec:CIVEW}

As we summarized in Table~\ref{tab:EVEWS}, the blue excess in J1443 can be almost explained by the contribution of the strong emission lines. 
This is also the case for J1202 by taking into account of the additional contribution of unobserved Ly$\alpha$ to $g$-band. 
On the other hand, the blue excess of the remaining two BluDOGs cannot be explained only by the contribution of BLR emission lines. 
Figure~\ref{fig:AFT} strongly suggests that a part of the excess flux comes from the continuum emission, which deviates at $\lesssim$7000\AA\ from the extrapolation of the power-law fit. 
These results demonstrate the complexity and diversity of BluDOGs; systematic exploration of a larger sample is required to statistically understand the origin of the blue excess.

\begin{figure}[ht!]
	%\centering
	        \hspace{-0.5cm}
		\includegraphics[width=9.0cm]{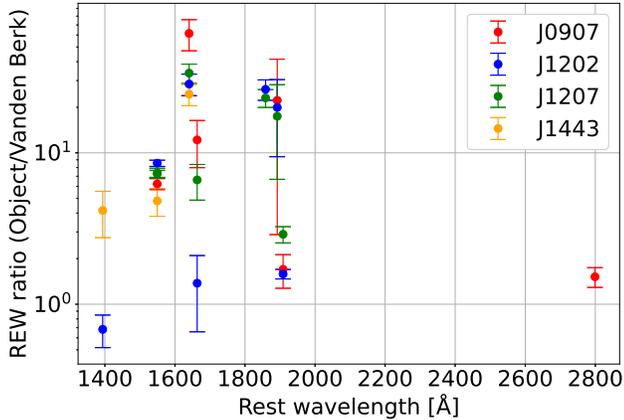} 
		%\vspace{30pt}
		\caption{Rest-frame EW ratio vs. rest-frame wavelength. REW ratios are defined as REWs of objects over REWs from the composite spectrum of SDSS type-1 quasar measured by \cite{2001AJ....122..549V}. The red, blue, green, and orange plots show the REW ratios of J0907, J1202, J1207, and J1443, respectively.}
		\label{fig:REWratio}
\end{figure}

\begin{figure}[ht!]
	%\centering
	        \hspace{-0.8cm}
		\includegraphics[width=10.cm]{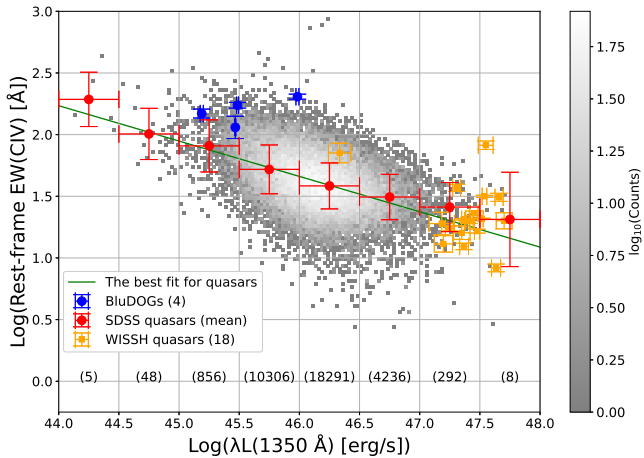} 
		%\vspace{30pt}
		\caption{Rest-frame EW of the C~{\sc iv} vs. the monochromatic luminosity at 1350 {\AA}. Blue and orange dots represent the four BluDOGs and WISSH quasars (\citealt{2018AandA...617A..81V}). The grey 2D histogram represents the number density of the SDSS quasars \citep{2011ApJS..194...45S}. The green line represents the linear fit to the distribution of the SDSS quasars. The red plots show the mean and standard deviation in luminosity bins with 0.5 dex width. The numbers of the SDSS quasars in the individual bins are shown at the bottom of the panel.}
		\label{fig:BaldwinEffect}
\end{figure}

Not only REW(C~{\sc iv}), but the REW of other BLR emission lines are also systematically larger than observed in typical type-1 quasars (see Tables~\ref{tab:dlineJ0907}--\ref{tab:dlineJ1443}, Figure~\ref{fig:REWratio}, and also Table~2 in \citealt{2001AJ....122..549V}). 
Such a trend may be explained if the observed BluDOGs have lower UV luminosity than typical quasars owing to the Baldwin effect (\citealt{1977ApJ...214..679B, 1990ApJ...357..338K, 2004MNRAS.350L..31B}), i.e., the negative correlation between the REWs and the continuum luminosities of quasars.
Figure~\ref{fig:BaldwinEffect} shows the four BluDOGs on the C~{\sc iv} REW vs. $\lambda L_{\lambda}$(1350{\AA}) diagram. 
Note that the REW of J1443 ($114\pm24$ {\AA}) is somewhat smaller than that of the remaining three BluDOGs ($148\pm12$, $203\pm10$ and $173\pm9$ \AA\ for J0907, J1202 and J1207, respectively; see Tables~\ref{tab:dlineJ0907}--\ref{tab:dlineJ1443}). This is partly because of an underestimation of the C~{\sc iv} flux caused by the absorption features.
The figure also shows SDSS type-1 quasars with reliable measurement of C~{\sc iv} REW (\verb|EWCIV/e_EWCIV > 5|) and without broad absorption lines (\verb|BAL < 1|) taken from \cite{2011ApJS..194...45S}. Since the Baldwin effect does not significantly depend on redshift (e.g., \citealt{2002MNRAS.337..275C, 2002ApJ...581..912D,2016ApJ...832..208N}), we do not adopt any redshift criterion to select the SDSS quasars so that a wide luminosity range is covered. We also use another comparison sample taken from the WISE/SDSS selected hyper-luminous quasar sample (WISSH; \citealt{2017AandA...598A.122B,2018AandA...617A..81V})\footnote{The C~{\sc iv} REW and C~{\sc iv} line luminosity of WISSH quasars are given by \cite{2018AandA...617A..81V}. To calculate $L_{\lambda}$(1350{\AA}) of WISSH quasars, we assume that the continuum spectrum of WISSH quasars is a power-law and adopt the following formula:
\begin{eqnarray}
   L_{\lambda} ({\rm 1350 \AA})&=&\frac{L^{\rm line} ({\rm C~{\sc IV}})}{REW ({\rm C~{\sc IV}})} \times \Big(\frac{1350}{1549}\Big)^{\alpha_{\lambda}},
\end{eqnarray}
where $L_{\lambda} ({\rm 1350 \AA})$, $L^{\rm line} ({\rm C~{\sc IV}})$, and $\alpha_{\lambda}$ are the monochromatic luminosity at 1350 {\AA}, the line luminosity of C~{\sc iv}, and power-law index, respectively. Here we adopt $\alpha_{\lambda} = -1.7$ \citep{2001AJ....122..549V} as the power-law index.
},
in order to add objects at the high-luminosity end.

Figure~\ref{fig:BaldwinEffect} clearly shows that the C~{\sc iv} REWs of BluDOGs are larger than the comparison samples at a given UV luminosity. 
The excess REW over the average relation of the Baldwin effect (shown with a green solid line in Figure~\ref{fig:BaldwinEffect}) for J0907, J1202, J1207, and J1443 are 0.29, 0.66, 0.44, and 0.26 dex, respectively.
This excess is larger than the scatter of the comparison samples (see red plots in Figure~\ref{fig:BaldwinEffect}). 
Therefore the large REW seen in the BluDOGs are not due to the Baldwin effect.

The averages and standard deviations of REW(C~{\sc iv}) for core ERQs and ERQ-like objects are $178\pm74$ and $86\pm45$ \AA, respectively \citep{2017MNRAS.464.3431H}.
The distributions of REW(C~{\sc iv}) and $(i-W3)_{\rm AB}$ color for BluDOGs are consistent with these of core ERQs although the most of core ERQs and ERQ-like objects do not show a blue-wing profile in C~{\sc iv} (Section \ref{subsec:SpecFeature}).
\cite{2017MNRAS.464.3431H} proposed a scenario that the large REW of ERQs are possibly due to the spatially extended geometry of BLRs caused by the powerful nuclear outflow. 
If the obscuration is heavier for the accretion disk than for the BLRs which have extended geometry, the continuum emission is more heavily extinct than the BLR emission lines and thus the observed-frame EW becomes larger.
Such a scenario may also apply to BluDOGs. 
Unfortunately it is not observationally feasible to confirm this idea by resolving the spatial structure of BLRs in ERQs or BluDOGs due to the required angular resolution, even with the JWST or exisiting ground-based interferometers. 
Without spatially resolving them, a possible approach is the velocity-resolved reverberation mapping of the geometry and kinematics of BLR clouds \citep[e.g.,][]{2004PASP..116..465H, 2009ApJ...704L..80D, 2013ApJ...779..110L, 2014A&A...566A.106K, 2014MNRAS.445.3073P}.

\subsection{Possible extreme accretion and the nature of BluDOGs}\label{subsec:DBHER}

\begin{figure}[ht!]
		%\hspace{-2.5mm}
		\includegraphics[width=8.8cm]{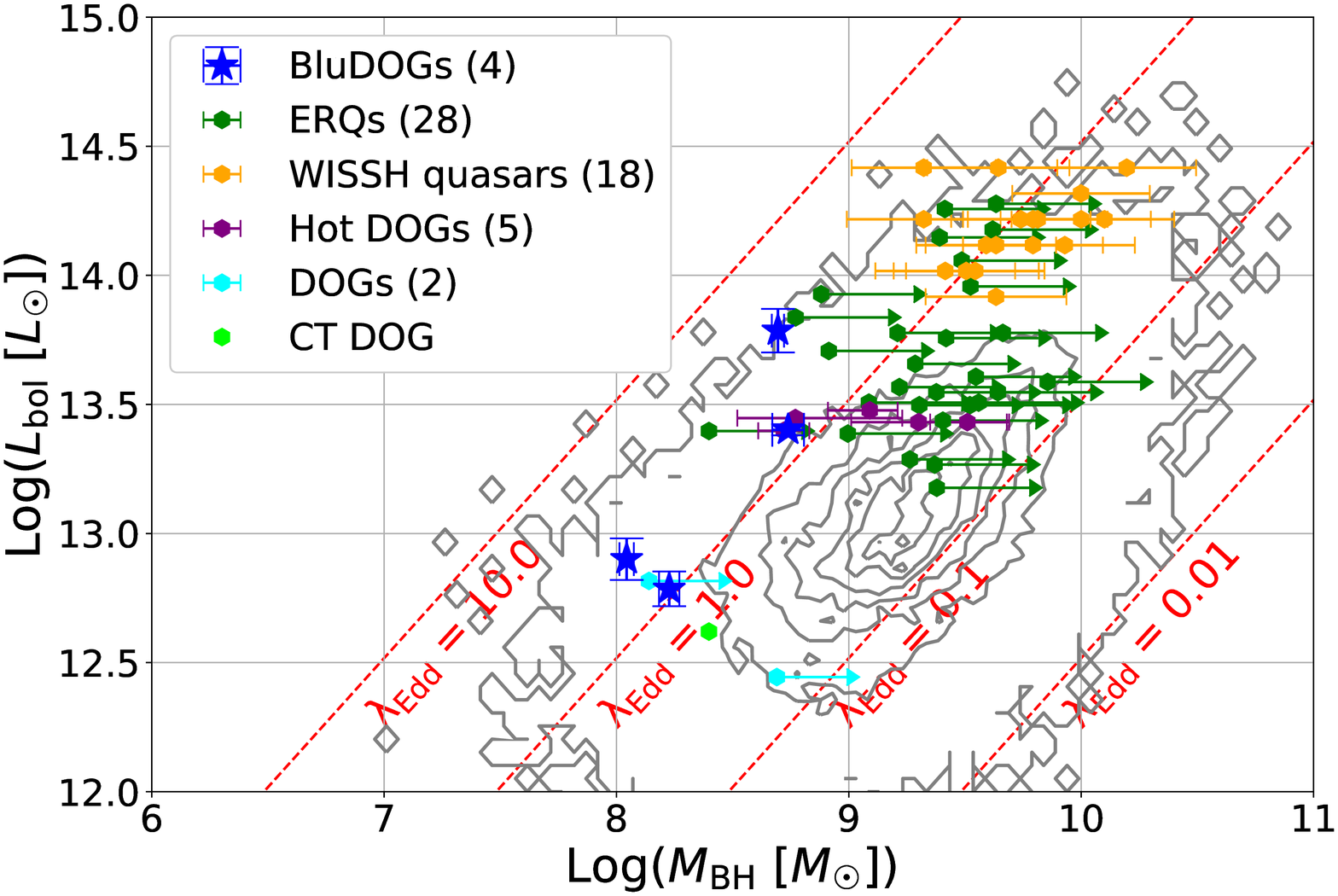} 
		%\vspace{30pt}
		\caption{The SMBH mass vs. the bolometric luminosity diagram. The filled-blue stars and gray contour denote the BluDOGs and SDSS quasars, while the filled hexagons with green, orange, purple, cyan, and light green colors denote ERQs \citep{2019MNRAS.488.4126P}, WISSH quasars \citep{2018AandA...617A..81V}, Hot DOGs \citep{2018ApJ...852...96W}, DOGs \citep{2011AJ....141..141M,2012AJ....143..125M}, and a CT DOG \citep{2020ApJ...888....8T}, respectively. The red dashed lines represent a constant Eddington ratio of $\lambda_{\rm Edd} =$ 0.01, 0.1, 1.0, and 10.0.}
		\label{fig:BHMvsLbol}
\end{figure}

To understand the nature of BluDOGs especially in the context of the major-merger scenario for the quasar evolution, we compare the SMBH accretion of the four BluDOGs with other AGN populations. 
Figure \ref{fig:BHMvsLbol} is a diagram of $L_{\rm bol}$ vs. the SMBH mass. 
As in Section~\ref{subsec:CIVEW}, SDSS quasars \citep{2011ApJS..194...45S} and WISSH quasars \citep{2018AandA...617A..81V} are used as comparison samples. 
For the SDSS quasars, we select only non-BAL quasars (\verb|BAL < 1|) with the uncertainty of $L_{\rm bol}$ and $M_{\rm BH}$ less than 0.5 dex (\verb|e_logBHCV < 0.5 & e_logLbol < 0.5|), and adopt the C~{\sc iv}-based SMBH mass for a fair comparison with those of the BluDOGs.
We also plot samples of 28 ERQs \citep{2019MNRAS.488.4126P}, 5 Hot DOGs \citep{2018ApJ...852...96W}, 2 power-law DOGs \citep{2011AJ....141..141M}, and 1 Compton-thick (CT) DOG \citep{2020ApJ...888....8T}. 
Hot DOGs are DOGs with a special color of {\it WISE} (very faint in the 3.4 $\mu$m and 4.5 $\mu$m bands, but bright in the longer bands; \citealt{2012ApJ...755..173E, 2012ApJ...756...96W}), while power-law DOGs are DOGs with a featureless power-law SED from optical to mid-IR \citep[e.g.,][]{2008ApJ...677..943D, 2012ApJ...744..150B, 2015PASJ...67...86T, 2019ApJ...876..132N}. 
The CT DOG was identified by {\it Nuclear Spectroscopic Telescope Array} \citep{2013ApJ...770..103H} from SDSS-{\it WISE} DOGs sample. 
All but the CT DOG have spectroscopic redshifts. 
The SMBH masses of the ERQs and the WISSH quasars are estimated from H$\beta$, while those of the Hot DOGs and the DOGs are estimated from H$\alpha$. 
The SMBH mass of the CT DOG was estimated by \cite{2020ApJ...888....8T} from the stellar mass by using an empirical relation between the stellar mass and SMBH mass \citep{2013ARA&A..51..511K}.
Since \cite{2019MNRAS.488.4126P} and \cite{2011AJ....141..141M} did not correct the absorption of dust, the $M_{\rm BH}$ of ERQs and DOGs are lower limits.

Figure~\ref{fig:BHMvsLbol} shows that the four BluDOGs are more luminous than the other AGN populations at a given SMBH mass, or equivalently, they have lower-mass SMBHs than the other AGN populations at a given bolometric luminosity. 
This suggests that the SMBH growth in the BluDOGs is more rapid than AGNs in comparison samples.
Indeed, the Eddington ratios ($\lambda_{\rm Edd}$) of J0907, J1202, J1207, and J1443 are $1.10\pm0.20$, $3.89\pm0.78$, $2.19\pm0.44$, and $1.40\pm0.23$, respectively (Table~\ref{tab:BHP}), with the average value of 2.26. 
In other words, the SMBHs in the BluDOGs are now in the stage of the Eddington-limit or super-Eddington accretion. 
Even if the intrinsic SMBH masses are lower than those estimated (Section \ref{subsec:M_SMBH_MASS}), the conclusion of this study remains qualitatively unchanged.
The higher Eddington ratios compared to other populations suggest that the SMBHs in BluDOGs are in the most rapidly evolving phase during the whole evolutionary history of SMBHs. 
In the gas-rich major merger scenario of \citet{2008ApJS..175..356H}, the peak of the AGN activity (i.e., the mass growth of SMBHs) corresponds to the transition phase from the optically thick to optically thin quasars, where the surrounding dust is blown out by the powerful AGN activity. 
Note that optically thick quasars in the major merger scenario should be recognized as type-2 quasars in optical (the BLR cannot be observed due to the heavy dust reddening). 
Since optically-thick quasars in the final stage of the evolution can be recognized as both type-1 and type-2 due to the orientation effect toward the dusty torus, the observed type-1 nature suggests the object is not in the early (optically-thick) stage in the major merger evolutionary scenario.
Preferentially in AGNs with high $\lambda_{\rm Edd}$, a blue-wing feature tends to be observed (e.g., \citealt{2005ApJ...618..601A, 2008ApJ...680..926K}).
The observed characteristics of the BluDOGs such as the type-1 nature and the blue-wing feature of the C~{\sc iv} velocity profile are consistent with the picture that BluDOGs are in such a peak stage of the SMBH evolution.

To discuss the evolutionary relation among populations of dusty galaxies (BluDOGs, core ERQs, and hot DOGs), we focus on $E(B-V)$ and $kt_{80}$.
$E(B-V)$ for BluDOGs, core ERQs \citep{2017MNRAS.464.3431H}, and hot DOGs \citep{2018ApJ...852...96W} are $0.273\pm0.049$, $0.242\pm0.127$, and $4.781\pm1.986$, respectively.
The $E(B-V)$ of the hot DOGs is significantly larger than that of the BluDOG and core ERQ samples, suggesting the hot DOGs are thought to be in a heavily obscured phase.
Since the $kt_{80}$ of BluDOGs is smaller than that of core ERQs (Section \ref{subsec:SpecFeature}), and the $kt_{80}$ of Mid-IR detected quasars is close to that of BluDOGs (Figure 1 of \citealt{2022MNRAS.511.3501M}), the BluDOG phase is thought to be close to the optically-thin quasar phase.
Therefore, it is suggested that the evolutionary path of various AGN populations is ``Hot DOGs -- core ERQs -- BluDOGs -- optically-thin quasars''.

For AGNs in general, the mass accretion efficiency ($\eta$) is defined as following: \begin{eqnarray}
	\eta &=& \frac{L_{\rm bol}}{\dot{M}c^2},
\end{eqnarray}
where $\dot{M}$ is the mass accretion rate. 
By multiplying the $\dot{M}$ and lifetime of BluDOGs ($t_{\rm life}$), we can roughly estimate the accreted mass ($M_{\rm Acc}$) in the BluDOGs phase.
\cite{2003PASJ...55..599B} estimated $\log\eta = -1.61$ of Seyfert 1 galaxies and Palomar-Green quasars by assuming that the geometrically-thin and optically-thick standard $\alpha$-prescription accretion disk model (\citealt{1973A&A....24..337S}).
By assuming $\log\eta = -1.61$ and $t_{\rm life} = 1$ Myr \citep{2019ApJ...876..132N}, the estimated $M_{\rm Acc}$ of J0907, J1202, J1207, and J1443 are about $1.68\times10^7$, $1.68\times10^8$, $2.19\times10^7$, and $6.93\times10^7\ M_\odot$.
The SMBH masses reached when the SMBH masses of the BluDOGs are increased by the observed mass accretion rate during the typical lifetime of BluDOGs ($M^{+}_{\rm BH} =  M_{\rm BH} + M_{\rm Acc}$) of J0907, J1202, J1207, and J1443 are $1.86\times10^8$, $6.63\times10^8$, $1.32\times10^8$, and $6.17\times10^8\ M_\odot$, respectively.
Therefore the SMBH mass of BluDOGs increases by $\sim$20\% during the short BluDOG phase, suggesting that BluDOGs are actually in a rapidly glowing phase.

\begin{figure}[ht!]
		%\hspace{-4.5mm}
		\includegraphics[width=9.0cm]{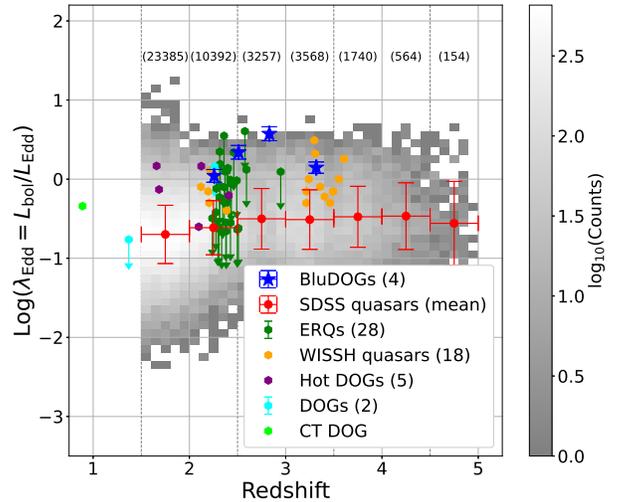} 
		%\vspace{30pt}
		\caption{The redshift vs. the Eddington ratio diagram. The filled-blue stars and hexagons are the same as in Figure \ref{fig:BHMvsLbol}. The gray 2D histogram represents the number density of SDSS quasars \citep{2011ApJS..194...45S}. The red plots show the mean and standard deviation of $\lambda_{\rm Edd}$ in redshift bins with the width of $\Delta z = 0.5$. The numbers shown at the upper part denote the numbers of SDSS quasars in the redshift bins.}
		\label{fig:LAMEDDvsREDSHIFT}
\end{figure}

Figure~\ref{fig:LAMEDDvsREDSHIFT} shows the Eddington ratios of various populations of AGNs as a function of redshift. 
The excess of $\lambda_{\rm Edd}$ of the four BluDOGs is more significant than the scatter of the $\lambda_{\rm Edd}$ distribution, suggesting that BluDOGs are a special class of AGNs that harbor SMBHs in the most actively evolving phase. Then, why such a class of AGNs is found only in a limited redshift range, $2.2 < z_{\rm sp} < 3.3$? 
A possible reason comes from their selection criteria, as briefly mentioned in Sec~\ref{subsec:SpecFeature}. 
Since the BluDOGs are selected by the blue excesses which are largely caused by the contribution of strong BLR emission lines, the resultant redshift distribution would be biased such that the blue bands contain strong emission lines.
It is also not clear whether the whole population of DOGs have systematically larger $\lambda_{\rm Edd}$ than ordinary type-1 quasars, due to the paucity of the spectroscopic data. In order to reveal the total picture of the dust-enshrouded evolution of SMBHs, more systematic spectroscopic observations for various populations of BluDOGs and DOGs are needed.

\section{CONCLUSION} \label{sec:conclusion}

We carried out spectroscopic observations of the four BluDOGs selected by \citet{2019ApJ...876..132N} using Subaru/FOCAS and VLT/FORS2. The analysis of the obtained spectroscopic data revealed the following spectroscopic properties of the BluDOGs:

\begin{enumerate}
\item The rest-frame UV spectra of the BluDOGs show broad ($\gtrsim$2000 ${\rm km\ s^{-1}}$) emission lines. This suggests that the BLRs of the BluDOGs are not completely obscured, albeit the very dusty nature inferred from their optical-IR SED.
\item The C~{\sc iv} lines of the BluDOGs show a significant blue wing, which is more prominent than in ordinary SDSS type-1 quasars. This suggests a presence of powerful nuclear outflow at the spatial scale of the BLR in the BluDOGs.
\item The REWs of their BLR lines are very large, REW(C~{\sc iv})$\sim$160 \AA, $\sim$7 times larger than the average of SDSS type-1 quasars. Such strong lines cause the flux excess of the two BluDOGs in the HSC $g$- and $r$-bands, while blue continuum emission also contributes the blue excess in the remaining two objects. The large REWs are not explained by the Baldwin effect. A possible origin is a powerful nuclear outflow in the BluDOGs causing a selective obscuration of the nuclear region, as suggested for ERQs.
\item The Eddington ratios of the BluDOGs are higher than 1.0 and are systematically larger than other AGN populations. The mass accretion onto the SMBH in BluDOGs is in the mode of the Eddington-limit or super-Eddington accretion. 
\end{enumerate}

All of the above results support the scenario that BluDOGs represent a population of AGNs in the transition phase from optically thick to optically thin quasars, i.e., in the blowing-out phase of the major-merger scenario for the SMBH evolution. 
The spectroscopic properties of the BluDOGs are similar to those of ERQs.
For further understandings of the complete picture, more systematic spectroscopic observations are crucial, not only of BluDOGs but also of the whole population of DOGs. 

%%%%%%%%%%%%%%%%%%%%%%%%%%%%%
\begin{acknowledgments}
The authors gratefully acknowledge the anonymous referee for a careful reading of the manuscript and very helpful comments. 
This study is based on data collected at Subaru Telescope, which is operated by the National Astronomical Observatory of Japan, and observations collected at the European Southern Observatory under a ESO programme 0102.B-0614(A).
We are honored and grateful for the opportunity of observing the Universe from Maunakea, which has the cultural, historical and natural significance in Hawaii.
%
%HSC
The Hyper Suprime-Cam (HSC) collaboration includes the astronomical communities of Japan and Taiwan, and Princeton University. 
The HSC instrumentation and software were developed by the National Astronomical Observatory of Japan (NAOJ), the Kavli Institute for the Physics and Mathematics of the Universe (Kavli IPMU), the University of Tokyo, the High Energy Accelerator Research Organization (KEK), the Academia Sinica Institute for Astronomy and Astrophysics in Taiwan (ASIAA), and Princeton University. 
Funding was contributed by the FIRST program from Japanese Cabinet Office, the Ministry of Education, Culture, Sports, Science and Technology (MEXT), the Japan Society for the Promotion of Science (JSPS), Japan Science and Technology Agency (JST), the Toray Science Foundation, NAOJ, Kavli IPMU, KEK, ASIAA, and Princeton University.
%HSC (LSST part)
This paper makes use of software developed for the Large Synoptic Survey Telescope. 
We thank the LSST Project for making their code available as free software at http://dm.lsstcorp.org.
%HSC (Pan-STARRS part)
The Pan-STARRS1 Surveys (PS1) have been made possible through contributions of the Institute for Astronomy, the University of Hawaii, the Pan-STARRS Project Office, the Max-Planck Society and its participating institutes, the Max Planck Institute for Astronomy, Heidelberg and the Max Planck Institute for Extraterrestrial Physics, Garching, The Johns Hopkins University, Durham University, the University of Edinburgh, Queen's University Belfast, the Harvard-Smithsonian Center for Astrophysics, the Las Cumbres Observatory Global Telescope Network Incorporated, the National Central University of Taiwan, the Space Telescope Science Institute, the National Aeronautics and Space Administration under Grant No. NNX08AR22G issued through the Planetary Science Division of the NASA Science Mission Directorate, the National Science Foundation under Grant No. AST-1238877, the University of Maryland, and Eotvos Lorand University (ELTE)
%
%VIKING
This publication has made use of data from the VIKING survey from VISTA at the ESO Paranal Observatory, program ID 179.A-2004. 
Data processing has been contributed by the VISTA Data Flow System at CASU, Cambridge and WFAU, Edinburgh. 
% WISE
This publication makes use of data products from the Wide-field Infrared Survey Explorer, which is a joint project of the University of California, Los Angeles, and the Jet Propulsion Laboratory/California Institute of Technology, funded by the National Aeronautics and Space Administration.
Herschel is an ESA space observatory with science instruments provided by European-led Principal Investigator consortia and with important participation from NASA.
This research made use of Astropy,\footnote{http://www.astropy.org} a community-developed core Python package for Astronomy \citep{2013A&A...558A..33A, 2018AJ....156..123A}.
We would like to thank Enago (www.enago.jp) for the English language review and Kohei Iwashita for helpful discussion.
This study was financially supported by the Japan Society for the Promotion of Science (JSPS) KAKENHI grant Nos. 19J10458, 21H01126 (A.N.), 16H03958, 17H01114, 19H00697, 20H01949 (T.N.), 18J01050, 19K14759, 22H01266 (Y. Toba), 20H01939 (K.I.), and 20K04014 (Y. Terashima).
\end{acknowledgments}

\software{Astropy, Astro-SCRAPPY, X-CIGALE, Numpy, Scipy.optimaize.curve\_fit, Recipe flexible execution workbench (Reflex)}
\facilities{Subaru (FOCAS, HSC), VLT (FORS2), VISTA, {\textit{WISE}}, and {\textit{Herschel}} (PACS, SPIRE)}

\bibliography{Ref}{}
\bibliographystyle{aasjournal}



\end{document}